\documentclass[12pt]{article}

\usepackage{graphicx}
\usepackage{amssymb}
\usepackage{amsthm}
\usepackage{booktabs}
\usepackage{rotating}
\usepackage{multirow}
\usepackage{eurosym}
\usepackage{amsfonts}
\usepackage{amsmath}
\usepackage{umoline}
\usepackage{caption}
\usepackage{subcaption}
\usepackage{color}
\usepackage{url}
\usepackage{authblk}
\usepackage[english]{babel}

\usepackage{pbox}
\usepackage{footnote}
\usepackage{multirow}
\usepackage{changepage}
\usepackage{tablefootnote}

\usepackage[margin = 2cm]{geometry}

\usepackage{hyperref}

\usepackage[natbibapa]{apacite}

\usepackage{rotating}
\usepackage{lscape}
\usepackage{float}
\restylefloat{table} 
\usepackage{booktabs}
\usepackage{textcomp}
\usepackage{siunitx}
\usepackage{multirow}
\usepackage{makecell}
\usepackage{longtable}
 
\begin{document}

\title{Super-App Behavioral Patterns in Credit Risk Models: Financial, Statistical and Regulatory Implications \footnote{\scriptsize NOTICE: this is the author's version of a work accepted for publication. Changes resulting from the publishing process, such as editing, corrections, structural formatting, and other quality control mechanisms may not be reflected in this document. Changes may have been made to this work since it was submitted for publication. Please cite this paper as follows: Luisa Roa, Alejandro Correa-Bahnsen, Gabriel Suarez, Fernando Cort\'{e}s-Tejada, Mar\'{i}a A. Luque, Cristi\'{a}n Bravo (2020). Super-App Behavioral Patterns in Credit Risk Models: Financial, Statistical and Regulatory Implications, Expert Systems with Applications: 114486. DOI: \protect\url{https://doi.org/10.1016/j.eswa.2020.114486}. \textcopyright CC-BY-NC-ND}}

\author[1]{Luisa Roa}
\author[1]{Alejandro Correa-Bahnsen \thanks{Corresponding author, alejandro.correa@rappi.com}}
\author[1]{Gabriel Suarez}
\author[2]{Fernando Cortés-Tejada}
\author[1]{Maria A. Luque}
\author[3]{Cristi\'{a}n Bravo}

\affil[1]{Rappi, Cl. 93 \#19-58, Bogotá, Colombia.}

\affil[2]{Pontificia Universidad Católica del Perú, Av. Universitaria 1801, San Miguel, Lima, Perú.}

\affil[3]{Department of Statistical and Actuarial Sciences, The University of Western Ontario, Western Science Centre, 1151 Richmond Street, London, ON, Canada.}

\date{}

\maketitle

\begin{abstract}
In this paper we present the impact of alternative data that originates from an app-based marketplace, in contrast to traditional bureau data, upon credit scoring models. These alternative data sources have shown themselves to be immensely powerful in predicting borrower behavior in segments traditionally underserved by banks and financial institutions. Our results, validated across two countries, show that these new sources of data are particularly useful for predicting financial behavior in low-wealth and young individuals, who are also the most likely to engage with alternative lenders. Furthermore, using the TreeSHAP method for Stochastic Gradient Boosting interpretation, our results also revealed interesting non-linear trends in the variables originating from the app, which would not normally be available to traditional banks. Our results represent an opportunity for technology companies to disrupt traditional banking by correctly identifying alternative data sources and handling this new information properly. At the same time alternative data must be carefully validated to overcome regulatory hurdles across diverse jurisdictions.
\end{abstract}

\begin{keywords}
Fintech;  Super-App; Credit Scoring; Financial Inclusion;  Alternative Data
\end{keywords}

\section{Introduction}

As technology companies have become more ubiquitous, their incursion into traditional lines of business has also become inevitable. In this context, the term \emph{Super-apps} has emerged, that is, mobile applications that in the same environment seek to satisfy different daily needs of consumers without requiring them to download another application. In other words \emph{Super-apps} play the role of a marketplace or ecosystem that hosts in itself different types of solutions, services and experiences that traditionally would only be found in an app specifically designed for it. Although the term \emph{Super-apps} began to develop mainly in Asia with giants like WeChat and Alipay, today more companies are looking to offer these all-in-one solutions to satisfy diverse customer needs. These offered solutions usually range from e-commerce to goods delivery, financial services and social networks. Furthermore these mobile applications target a large number of everyday needs of consumers, and in the financial sphere they offer services traditionally available through banks; while simultaneously taking advantage of its ability to generate a large amount of diverse data through each of the multiple daily interactions of its users on different fronts, data that has never before been available to traditional financial institutions. Consequently, as the use of these Super Apps increases, it becomes more and more relevant to assess the scientific questions that arise about how relevant these variables are and how these data sources can be used to improve the disruptive potential of these growing companies in the financial sector market, along with the regulatory implications of doing so.

The capacity of super-apps to expand far beyond their current spheres and become important financial technology (fintech) companies is becoming more common as they feature products and services that revolutionize not only online commerce but traditional financial services. Super-apps provide an ecosystem of services on one platform, thus, allowing their makers to cross-sell and improve user loyalty \citep{DBS2019}. Part of a super-app’s diversity is attributable to its “mini-program” function, which allows it to have the same functionalities as a specialized app directly within the super-app interface. Some examples of this are provided in Figure 1. In each of the functionalities of the super-apps, the data and information provided by the users’ selections are generated, and these data become distinctive attributes of the users’ behavior.

\begin{figure}[tbh!]
    \centerline{\includegraphics[scale = 0.30]{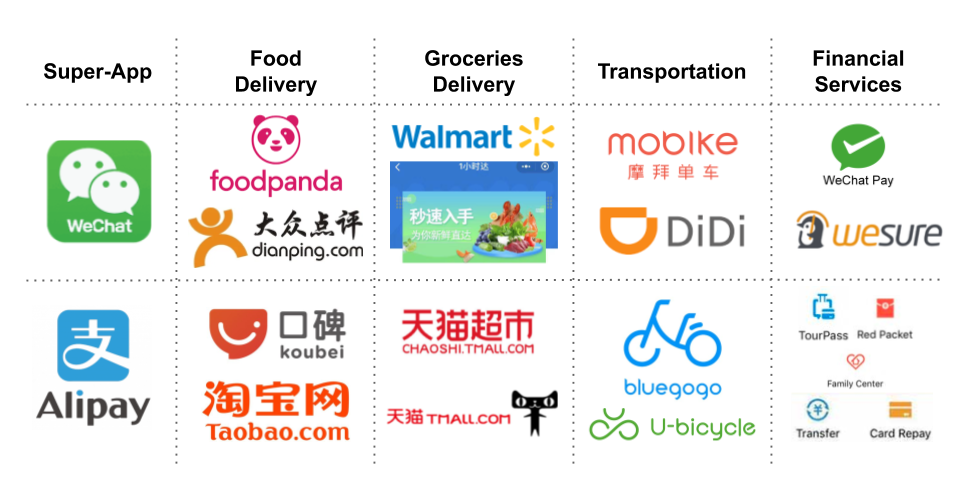}}
    \caption{Super-apps functionalities and mini-programs}
    \label{Super-apps}
\end{figure}

Although entering the financial market represents a great challenge, super-apps boast a competitive advantage over traditional banking as they possess data generated by the users of the platforms as well as the transactional data (once they have launched financial services) common to banking. The super-app companies can also serve as agents of financial inclusion by using their transactional and behavioral user data to assess and create tailored financial services that are targeted at these underserved segments. This is particularly important in emergent economies, as their household financial inclusion levels hardly break 50\% in some places, such as Latin America \citep{dabla-norris2015}. In terms of credit risk, these new data sources are known as  \emph{alternative data} \citep{Siddiqi2017} as they are derived from sources other than traditional banking and financial behavior. A considerable amount of evidence has mounted with regard to the potential for financial inclusion for these sources of data \citep{Bravo_2013,gool2012,Oskarsdottir2019}, particularly in countries with a large proportion of young and/or unbanked individuals where the super-app may achieve deeper market penetration compared with the traditional financial system.

A clear example of successful financial inclusion is Ant Financial, which has taken advantage of big data analytics, machine learning systems and deep learning to develop a wide range of intelligent products and services such as insurance, micro loans, payments, risk management services and other, which focus upon the needs of individuals and small businesses. The world's biggest unicorn began as an Alibaba strategy in 2004 to increase trust in the company among online buyers and sellers and has grown to become a world leader in financial innovation and risk management \citep{Sun2017}. For credit analysis, Ant Financial provides a score based on personal financial accounts from Ant Financial Services, social network and e-business information from the Alibaba Group platform and public utilities information \citep{Zhang2016}. The creditworthiness assessment by means of Ant Financial’s own scoring allows them to provide financial services to all Alibaba users, including non-users of the traditional financial system.  Similarly, the Fintech Lufax from the Ping An financial group offers more than 5,000 financial services to market segments that previously had no access to such services until this users transitioned to a technology company by connecting borrowers with investors \citep{osterwalder2020}. To understand the financial preferences of its users more deeply, Lufax generates models based upon natural language learning and user behavior data to identify and predict the needs of each user. Therefore, in each moment of the user’s life cycle the right products are offered, and the matches effected between borrowers and investors are more accurate and efficient \citep{waic2019}.

Alternative data from super-apps seem to promise the additional benefit of enhancing traditional credit score models; hence, we explore this in the paper and attempt to answer the following research questions:
\begin{enumerate}
  \item Is there an additional predictive value when considering the variables provided by a super-app?
  \item Is the value added by the variables of a super-app significant?
  \item What new behaviors do these variables reveal and how do they differ from traditional banking resources?
  \item What are the consequences of using super-app data for lenders, users and regulators?
\end{enumerate}

The rest of the paper proceeds as follows. Section 2 presents a review of the credit scoring and bank regulation literature related to fintech. Sections 3 and 4 describe the methodology and the experimental setup used within the research. In Section 5, the results are presented along with a discussion of their implications. Conclusions are drawn in Section 6 along  with the possibilities for future work on alternative data models for super-apps.

\section{Background}

\subsection{Credit Scoring}

To strategically manage risk, financial institutions assign each of the customers a credit score according to their estimated individual probability of a user defaulting on their obligations. This practice allows companies to define the level of risk at which they are willing to operate and, therefore, minimize the potential losses to which they may be exposed. The objective of this credit score for each client is to classify whether they are more or less likely to default on their financial obligations and to assess whether or not they will be approved for potential credit under  the risk levels accepted by the institution \citep{Lawrence2012}. Typically, different financial companies around the world have addressed this classification problem through standard cost-insensitive binary classification algorithms, such as logistic regression, neural networks, discriminant analysis, genetic programming, and decision trees, among others \citep{LESSMANN2015124}.
  
Formally, a credit score is a statistical model that allows the estimation of the probability $\hat p_i=P(y_i=1|\mathbf{x}_i)$ of a customer $i$ defaulting upon a contracted debt. Additionally, since the objective of credit scoring is to estimate a classifier $c_i$ to decide whether or not to grant a loan to a customer $i$, a threshold $t$ is defined such that if $\hat p_i < t $, then the loan is granted, that is, $c_i(t)=0$, and denied otherwise, that is, $c_i(t)=1$ \citep{thomas2017}.
  
\subsubsection{Credit Bureau Features}\label{traditional_features}

Some of the most commonly used variables around the world in the conformation of these models are the scores generated by a credit bureau or consumer reporting agency, that is, companies dedicated to collecting data upon individuals throughout their financial lives, who then makes this information available this information to the market through credit reports for a possible lender to purchase \citep{Hurley2016}. To this end, the credit bureau examines how individuals have behaved with the financial companies with which they have interacted and generates a quantitative score from this information, which is commonly used as an indicator for lending companies to assess the probability of the individual defaulting. In many countries, the bureau scores are synonymous with a credit score but most modern banks use their own implementations, which only utilize the bureau score as an input. Among other things, the variables that constitute the financial report are the number of credits in history, the type of credits acquired, the use of these credits or how many of them are available, possible debts,payment defaults within a history, and bankruptcies or late payments. These variables can be used as they are, or the score itself can be used as a first variable as in the case of this paper.

\subsection{Alternative Data and Fintech}

Fintech has assumed great importance in recent years functioning in the financial sector to provide online solutions for payments, transfers, investments and lending, among other services. Since 2010, more than U.S.\$50 billion has been invested in 2,500 fintechs worldwide \citep{Sy2019} and it is estimated that by 2025 global fintech market size will be  U.S.\$124.3 billion with a compound annual growth rate of 23.84\% \citep{valuatesreports2019}. The growth of fintech can be partly explained by the advantages that it offers compared to traditional financial entities, which include lower costs, greater speed, lower rates and, in particular, ease of access to a larger pool of products \citep{Philippon2019}.

In 2017, the number of adults worldwide who did not have a basic transaction account was 1.7 billion \citep{Asli2018}. These individuals are essentially invisible to traditional banking because of they lack a banking history. For this population, fintechs plays a fundamental role as it strengthens financial development by providing services that challenge the traditional structures within the financial services value chain \citep{Sy2019}. In this regard, alternative data can be essential to providing insights into the financial behavior of unbanked consumers, as it can be used to strengthen current credit scores and can allow for greater precision than that provided by the scores created using traditional credit assessments.

To ascertain the financial behavior of unbanked customers and obtain a proxy for a credit history, non-traditional data grouped into non-financial payments, individual behavior and data ingestion can be used \citep{Aitken2017}. The first group, non-financial payments, includes data recording whether a consumer has made the payments that they have promised to make, for instance, payment records of basic utilities such as gas, water and electricity, telecommunications services and rent. This information also includes the individual’s asset record \citep{Carroll2017}. These non-financial payments are conceptually related to traditional data given that they measure the ability to pay in different environments and  they are based upon the same principle of whether an individual is credit worthy or not. These variables have previously been shown to be powerful predictors of default in unbanked segments in the past \citep{Bravo_2013}. The second group consists of information that diverges from financial data and measures individual behaviors such as consumption patterns, criminal records, traffic violations, employment history and address changes. These can be converted into insights into reliability and creditworthiness with careful consideration of the ethical and regulatory implications. Variables such as stable work patterns, for example, can signal a lower defaulter behavior \citep{Aitken2017}. Finally, data ingestion attempts to identify the unbanked through the data trails generated in applications, web pages, social media, call log, emails and messages. The importance of data ingestion is that it is the most inclusive of all the non-traditional data forms since it is easier to obtain access to a mobile phone than to a financial service. Consequently, with each online movement new demographic, geographic, financial and social data are generated that strengthen financial inclusion and credit scores.

Different authors have used alternative data in credit prediction models, thereby, seeking to improve the model performance and demonstrate that non-traditional data are valuable to the financial sector. First, \citet{ZhangHengyue2016} presented a credit score model that merged traditional information with information from the social networks of the users of a peer-to-peer platform. They found that of the six most predictive attributes of default, two were social network information and these also outperformed traditional credit ratings. Moreover, \citet{Oskarsdottir2019} developed a credit scoring model based upon mobile phone data, call networks and default propagation and found that the models that included call data performed better than traditional credit scoring models, both statistically and financially. The study showed that using the alternative information, even by itself, could lead to predictive results as precise as those provided by traditional information. In addition to call networks and social media data, the use of digital footprints has also been shown to provide complementary information to traditional credit score data. \citet{Berg2019} demonstrated how behavior on a website improves the default prediction of those individuals who are already in the financial system and enables the reduction of informational asymmetry when the bureau score is not available by predicting only with digital footprint variables and doing so effectively.

Nonetheless, the use of this alternative information leads to some regulatory challenges as it is necessary to have white box models that facilitate interpretation and the variables extracted from the different alternative sources must be accurate, predictive and transparent \citep{Wu2019}.

\subsection{Banking Regulation}

Banking regulation is fundamentally related to credit scoring. The estimation of the probability of default (PD) is a function (usually a segmentation) of the score, adjusted by microeconomic factors \citep{baesens2016}. This means the development and deployment of credit scores is highly regulated and must pass the stringent controls imposed by local banking regulators. Fintechs challenge the traditional methods used by banks through the design and implementation of machine learning models that, seem to have greater predictive and classification power but may lack interpretability \citep{lime}. Furthermore, these complex algorithms may unintentionally incorporate variables that are proxies for sensitive consumer attributes \citep{Hurley2016}. It is, therefore, mandatory for regulators to mitigate the potential risks of these new approaches, and in this way ensure that the scoring decisions are as accurate as possible but also as unbiased, transparent and fair as possible \citep{BCBSFintech2018}. We will comment on the regulatory implications of our findings in Section~\ref{sec:Regulatory}.

\section{Methodology}

This paper contributes to the literature by investigating the use of transactional data for credit scoring in a fintech context, explaining the new behaviors have for lending. In this section, the methodology proposed for combining and extracting valuable features from a super-app is discussed. Moreover, the financial evaluation measure used in the experiments is presented.

\subsection{Users interactions with super apps}

Users interact with super-apps in quite different ways, every movement of the user in the application generates features that might be useful when creating a credit scorecard. The transactions carried out by users contain a record of the different characteristics of the users and their behavior. Each transaction generates data such as the transaction amount, the type of transaction, the payment method, date, and the type of store, among others. Given that these super-apps fulfill numerous functions, many possible features can be extracted from the most popular and common functionalities they possess, such as being used for food and grocery deliveries, or transportation or financial services engagement.

App-based features, as discussed above, have demonstrated to add value to credit risk assessment in other studies. However, previous works have been limited to consider users' characteristics and overall behavior within websites or apps, including data such as device, operating system, check-out time, email veracity and social network \citep{Berg2019,Carroll2017,guo2016}. Although these features are valuable and some are also considered in this study, we go beyond these analyses. The novelty of the features presented in our work lies in the collection and aggregation of transactional, consumption and payment variables segmented by the different functionalities that a super-app provides. 

Some examples of the features we collected for this study can be observed in Table~\ref{tab:Features}, where Sum, Pct, Avg, Count, Max presents for the aggregate of the specific variable through time, the percentage of consumption of that variable when compared to total consumption, the average value that the variable has had over time, the number of occurrences of that variable and the maximum value the variable has obtained in time, respectively. It should be noted that not all of these features must necessarily be included for the formation of the credit scorecard, as the variables must be carefully chosen in order to avoid building discriminatory or subjective scores.

\textbf{Generic features:} These refer to the demographic qualities of the user. These data may include attributes such as age, gender, place of residence, brand of cell phone as well as social characteristics such as income. These features provide an overview of the type of user and are mainly used to understand segments within the application to adapt offers and campaigns.

\textbf{Delivery:} This functionality includes all the services related to the purchase and delivery of food, groceries, technology, clothing, pharmaceutical products and others. Variables that can be created from this type of service allow an understanding to be formed of consumer consumption patterns, user preferences and how  users make use of different types of stores.

\textbf{Transportation:} This considers the data generated by the scooter, bike and ride sharing system operators such as Didi, Uber, Lime, Mobike and others. The features that can be extracted from this functionality provide information about the movements of people in a geographic area such as mobility patterns and the most frequently used transportation method.

\textbf{Financial services:} This last functionality collects financial services or products delivered via technology ranging from e-wallets and digital cards to loan services, on- and offline payments, and money transfers. These financial services allow the features associated with the number of products and the users' financial behaviors to be defined.

\renewcommand\theadalign{bc}
\renewcommand\theadfont{\bfseries}
\renewcommand\theadgape{\Gape[4pt]}
\renewcommand\cellgape{\Gape[4pt]}

 \renewcommand{\baselinestretch}{0.9}

 \renewcommand{\arraystretch}{0.9}
 \begin{table}[tbp!]
  \caption{Feature types with examples}
 \centering
     \begin{tabular}{lp{10cm}}
     \toprule
     Feature type                        &   Examples         \\ 
     \midrule
     \multirow{10}{8em}{Generic features}   & 1. Gender        \\ 
                                            & 2. Age range and age in the app (tenure)\\
                                            & 3. Country/city of residence        \\
                                            & 4. Most used address     \\ 
                                            & 5. Number of different addresses     \\ 
                                            & 6. Preferred payment method   \\ 
                                            & 7. Number of registered credit cards (national/international)    \\ 
                                            & 8. Number of registered credit card brands   \\
                                            & 9. Phone brand/operating system   \\
                                            & 10. Number of different phones used   \\
     \midrule
     \multirow{15}{8em}{Delivery by vertical (delivery, groceries, pharmacies and others) }      &1.  Sum/Pct/Avg/Count/Max of total orders \\
                                            &2.  Sum/Pct/Avg/Count/Max of approved orders\\
                                            &3.  Sum/Pct/Avg/Count/Max of orders value      \\
                                            &4.  Sum/Pct/Avg/Count/Max of total number of cancelled orders (By user/Payment Error/Fraud) \\
                                            &5. Sum/Pct/Avg/Count/Max of total refund\\
                                            &6. Sum/Pct/Avg/Count/Max of payment method used\\
                                            &7. Sum/Pct/Avg/Count/Max of total discount\\
                                            &8. Sum/Pct/Avg/Count/Max of consumption in a certain vertical\\
                                            &9. Sum/Pct/Avg/Count/Max offered tip \\
                                            &10. Sum/Pct/Avg/Count/Max of value spent in a certain type of store\\
                                            &11 Avg/Count products per order\\
                                            &12 .Sum/Count of consumption in top store\\
                                            &13.Count of different stores in which a user purchases\\
                                            &14.Period of time when orders are placed \\
                                            &15.Store where a user purchases the most\\

     \midrule
      \multirow{7}{8em}{Transportation}       &1. Count of rides \\
                                              &2. Sum/Count/Avg/Max of travel time\\
                                              &3. Count of different departures locations\\
                                              &4. Count of different destinations\\
                                              &5. Most frequented destination\\
                                              &6. Count of sectors in the city within which a user has moved\\
                                              &7. Favorite transportation vehicle\\
     \midrule
     \multirow{8}{10em}{Financial Services} &1. Count of financial services \\
                                              &2. Sum/Pct/Avg/Count/Max of debit transactions\\
                                              &3. Sum/Pct/Avg/Count/Max of credit transactions  \\
                                              &4. Sum/Pct/Avg/Count/Max of total amount traded on debit cards\\
                                              &5. Sum/Pct/Avg/Count/Max of total amount traded on credit cards\\
                                              &6. Sum/Avg/Count/Max amount of transfers\\
                                              &7. Number of people to whom the user made transfers\\
                                              &8. Whether the user makes cash withdrawals\\

      \bottomrule
     \end{tabular}
 \label{tab:Features}
 \end{table}

\subsubsection{Combining the features}
\begin{figure}[h]
    \centerline{\includegraphics[scale = 0.50]{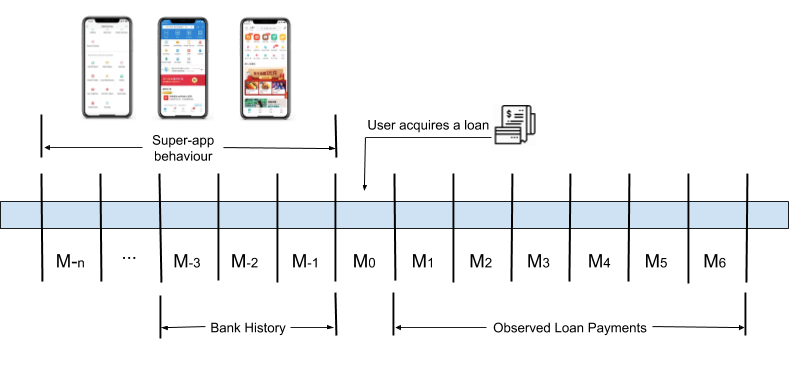}}
    \caption{Experimental Setup}
    \label{fig:set_up}
\end{figure}

Data trails generated by the users of super-apps in their different functionalities become important as a mean of supplying behavioral and purchasing patterns. For a user who retains a financial service, the relationship between traditional features, such as a bank history, and super-app features represents an additional value for the credit evaluation. Figure \ref{fig:set_up} presents how the historical behavior on the super-app and the bank history collect information to predict payment defaults for any user who has received a loan. In the months after the user receives the credit, the real value of whether or not the user has a late payment allows the user to be classified as a defaulter or a non-defaulter.

\subsection{Financial Evaluation Measure}\label{sec_savings}

Traditional measures to evaluate credit scoring models include the area under the receiver operating characteristic curve (AUC), the Brier score, the Kolmogorov-Smirnov (K-S) statistic, the F1-Score, and the misclassification rate \citep{LESSMANN2015124}. Nevertheless, none of these measures takes into account the business and financial realities that take place in lending. The costs incurred by the financial institution to acquire customers, or the profit expected from a particular client, are not considered in the evaluation of the different models \citep{CorreaBahnsen2015}.
Recent approaches have included the Expected Maximum Profit (EMP) measure \citep{VERBRAKEN2014505} and the example-dependent cost-sensitive approach (EDCS) for credit scoring \citep{CorreaBahnsen2014b}, both of which we used in this work. The former uses a distributional estimation for the losses, while the latter uses a fixed loss value where the entire distribution is located in a single point. Thus, the EMP measure will give a more forward-looking, albeit less certain, view of the expected profit, while the ECDS approach will give a more realistic expectation of the short-term performance of the model.

The EMP approach developed by Verbraken et al. \citet{VERBRAKEN2014505} is a profit-based classification performance measure, where the expected profits and losses of a granted loan are considered in order to assets a business-focused performance. EMP assumes constant ROI and a Loss Given Default (LGD) distribution with two point masses $\lambda=0$ with a probability $p_0$ of complete recovery and $\lambda=1$ with probability $p_1$ of complete loss. Thus, EMP is defined as 
$$EMP = \int_{0}^{1} P(T(\theta);\lambda,ROI)\cdot h(\lambda) d\lambda $$
where $P(T(\theta);\lambda,ROI)$ is the profit of classifying a borrower and $h(\lambda)$ the probability of $\lambda$ that follows a uniform distribution between 0 and 1. 

\begin{table}[bh!]
  \centering
\caption{Credit scoring example-dependent cost matrix}
\label{tab:c_mat}
\begin{tabular}{c|cc}
		\multicolumn{1}{c|}{}  & Actual Positive& Actual Negative \\
		\multicolumn{1}{c|}{} & $y_i=1$& $y_i=0$ \\
		\midrule
		Predicted Positive 		& \multirow{ 2}{*}{$C_{TP_i}=0$} & \multirow{ 
		2}{*}{$C_{FP_i}=r_i+C^a_{FP}$} 
		\\
		$c_i=1$ & &\\
		Predicted Negative  	& \multirow{ 2}{*}{$C_{FN_i}=Cl_i \cdot L_{gd}$} & \multirow{ 
		2}{*}{$C_{TN_i}=0$} \\
		$c_i=0$ & &\\
	\end{tabular}
\end{table}
	  
As for the example-dependent cost-sensitive approach \citep{CorreaBahnsen2014b}, in Table \ref{tab:c_mat}, the credit scoring cost 
matrix  is shown. Initially, the costs of a correct classification, $C_{TP_i}$ and $C_{TN_i}$, are zero for all customers, $i$. Then, $C_{FN_i}$ reflects the incurred losses if the customer $i$ defaults, which is proportional to ther is credit line $Cl_i$ and the cost of a false positive $C_{FP_i}$ as the sum of two real financial costs $r_i$ and $C^a_{FP}$, where $r_i$ is the loss in profit through rejecting what someone who would have been a good customer \citep{Nayak1997}. 
  
Finally, the cost improvement can be expressed as the cost savings as compared with $Cost_l$. 
  $$    Savings = \frac{ Cost_l - Cost}   {Cost_l},$$  
where $Cost$ is calculated as 
  $$   Cost = \sum (1-c_i)*y_i*C_{FN_i} + (1-y_i)*c_i*C_{FP_i},$$
and $Cost_l$ is the cost of the cost-less class \citep{CorreaBahnsen2015}.

\section{Experimental Setup}

\subsection{Data}
Our dataset consisted of the transactional information of users within a super-app for two different Latin American countries, labeled as Country A and B. In the first country, a sample of 50,000 users was studied, while 30,000 users were analyzed for the second.  For each user, we had access to all their transactional data within the super-app, which included orders placed to more than  15,000 restaurants and 2,000 grocery stores. In addition, we had access to several observations regarding each of the users, such as the location in which they requested their orders, the device and the operating system through which the user placed the orders and the data regarding the payment method used, including – when applicable – their credit card information. Moreover, we also had access to data that made it possible to determine consumption patterns and construct variables that characterized their financial behavior.

\subsection{Setup} 

Seeking to understand whether default prediction can be improved for certain populations, three segments were defined to divide the population into a sample with a high segment value and another with a low segment value. The first segment divided the population by device score as this a variable that allows an approximation of the economic potential of an individual \citep{Sundsoy2016}, while the second segment was intended to be a more robust approximation of the economic potential and to be associated with the behavior in the super-app, which we named Wealth Score. Finally, the last segment separated the population by a super-app user segmentation \citep[Recency, Frequency and Monetary Value;][]{fader2005} based on the recency since the user made their last purchase, the frequency with which they placed orders and the average amount spent. This segmentation has proven to be a valuable variable for other models developed internally within the super-app and in many applications. For each proposed segmentation and for the dataset without segmentation, three models were created: one that only contemplated the Bureau score, other taking into account just super-app features, and another that considered both types of features. The number of observations and the default rate for each country and segmentation are shown in Table~\ref{data_info}.

\renewcommand{\baselinestretch}{1}
\renewcommand{\arraystretch}{1.1}
\begin{table}[t!]
\centering
\caption{Dataset information}
    \begin{tabular}{lrrrr}
    \toprule
    & \multicolumn{2}{c}{Country A} & \multicolumn{2}{c}{Country B} \\ \cmidrule{2-3} \cmidrule{4-5}
    Model               & \multicolumn{1}{l}{Size}   & Default Rate    & \multicolumn{1}{l}{Size}      & Default Rate     \\
    \midrule
    No Segments         & 50,000  & 5.00\%          & 30,000    & 9.00\%            \\
    Low Device Score    & 29,627  & 5.52\%          & 14,548     & 8.51\%            \\
    High Device Score   & 20,373  & 4.24\%          & 15,452     & 9.46\%            \\
    Low Wealth Score    & 27,570  & 5.88\%          & 19,664     & 9.04\%            \\
    High Wealth Score   & 22,430  & 3.92\%          & 10336     & 8.92\%            \\
    Low RFM             & 26,479  & 5.87\%          & 15,998     & 9.40\%            \\
    High RFM            & 23,521  & 4.02\%          & 14,002     & 8.54\%            \\
    \bottomrule
    \end{tabular}
	\label{data_info}
\end{table}

An XGBoost classifier was implemented as it has demonstrated its superior performance over models such as neural network, decision tree, support vector machines and bagging-NN with regard to structured data \citep{Xia2017, Salvaire2019}. The final model performance was evaluated using a randomized bootstrap of 50 iterations on the databases with a data proportion of 70\% to train and the remaining 30\% to test in each iteration.

Model performance was measured by using the area under AUC and the KS measure.  The AUC captures the trade-off between true and false positives at various discrimination thresholds, while the KS statistic measures the degree of separation between two cumulative distributions, specifically the maximum distance for all classification thresholds between the true and false positive rate curves.

In addition, the financial performance of the models was evaluated with the EMP and estimated financial savings as described in Section~\ref{sec_savings}. The parameters required to estimate the EMP and savings are shown in Table~\ref{tab:EMP} and Table~\ref{tab:savings}.

These measures allowed us to assess the discriminatory ability for defaulters, and the average performance and the maximum performance for the most optimistic case. In addition, statistical tests were performed to establish whether there was a significant difference in the classification performances of any of the models. This was in order to compare the performances of the different segments and identify in which population, with a common characteristic, the super-app variables had a representative contribution.

\begin{table}[t!]
    \centering
	\caption{Parameters to estimate the EMP}
    \begin{tabular}{lc}
    \toprule
    Parameter & Value \\
    \midrule
    Complete recovery ($P_0$) & 20\% \\
    Complete loss ($P_1$) & 60\% \\
    ROI & 75\% \\
    \bottomrule
    \end{tabular}
	\label{tab:EMP}
\end{table}

\begin{table}[t!]
    \centering
	\caption{Parameters to estimate the financial savings}
    \begin{tabular}{lc}
    \toprule
    Parameter & Value \\
    \midrule
    Interest rate ($int_r$) & 40\% \\
    Cost of funds ($int_{cf}$) & 10\% \\
    Loss given default ($L_{gd}$) & 75\% \\
    \bottomrule
    \end{tabular}
	\label{tab:savings}
\end{table}

\section{Results and Discussion}

In this section we present the experimental results. First, the statistical performance results are described, followed by the financial performance results for each model. Finally, the regulatory implications of the results are discussed.

\subsection{Statistical model performance}

\begin{figure}[tbp!]
\begin{subfigure}{\textwidth}
    \centerline{\includegraphics[scale = 0.40]{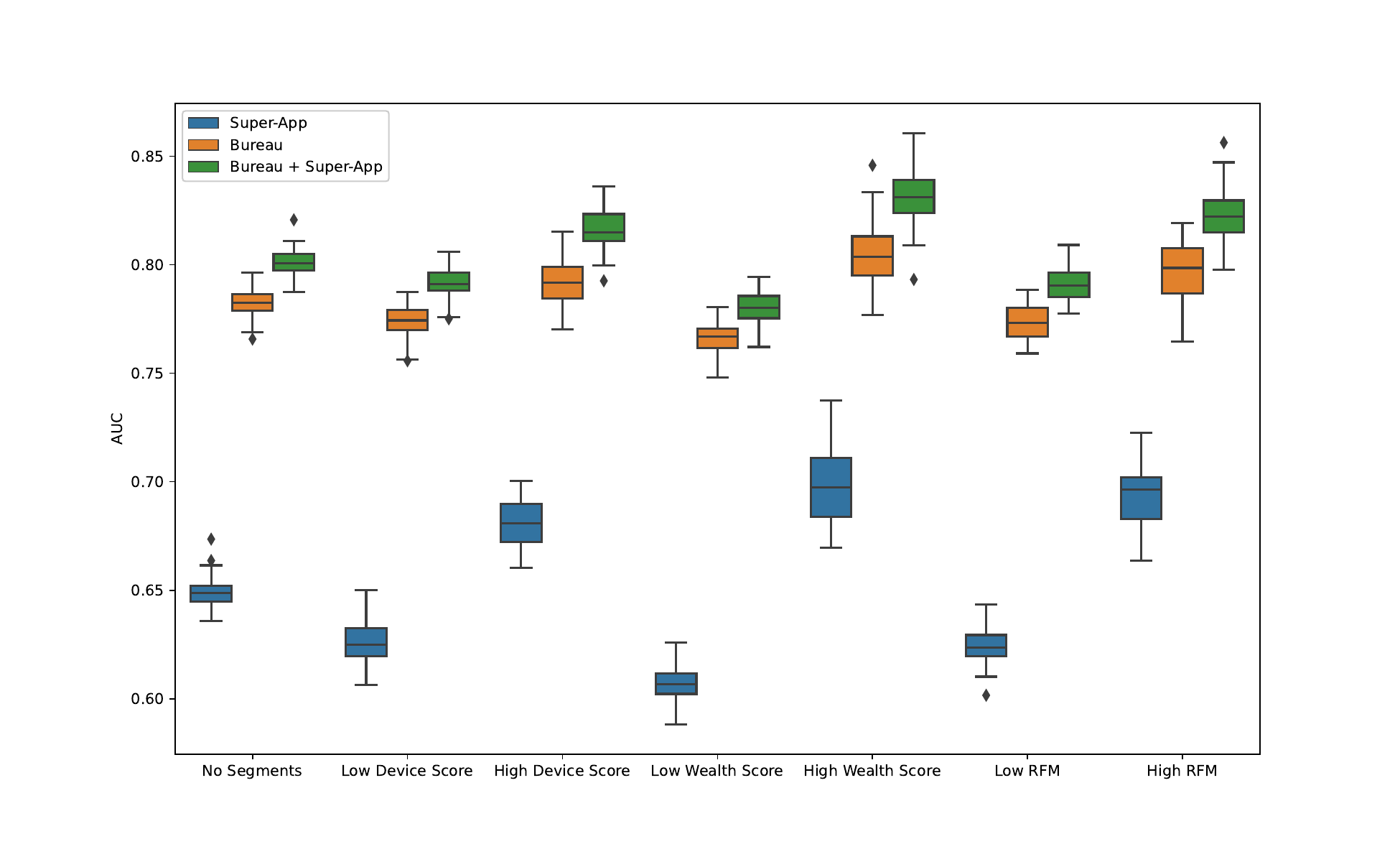}}
    \caption{Country A}
    \label{model_perf_auc_a}
\end{subfigure}

\begin{subfigure}{\textwidth}
    \centerline{\includegraphics[scale = 0.40]{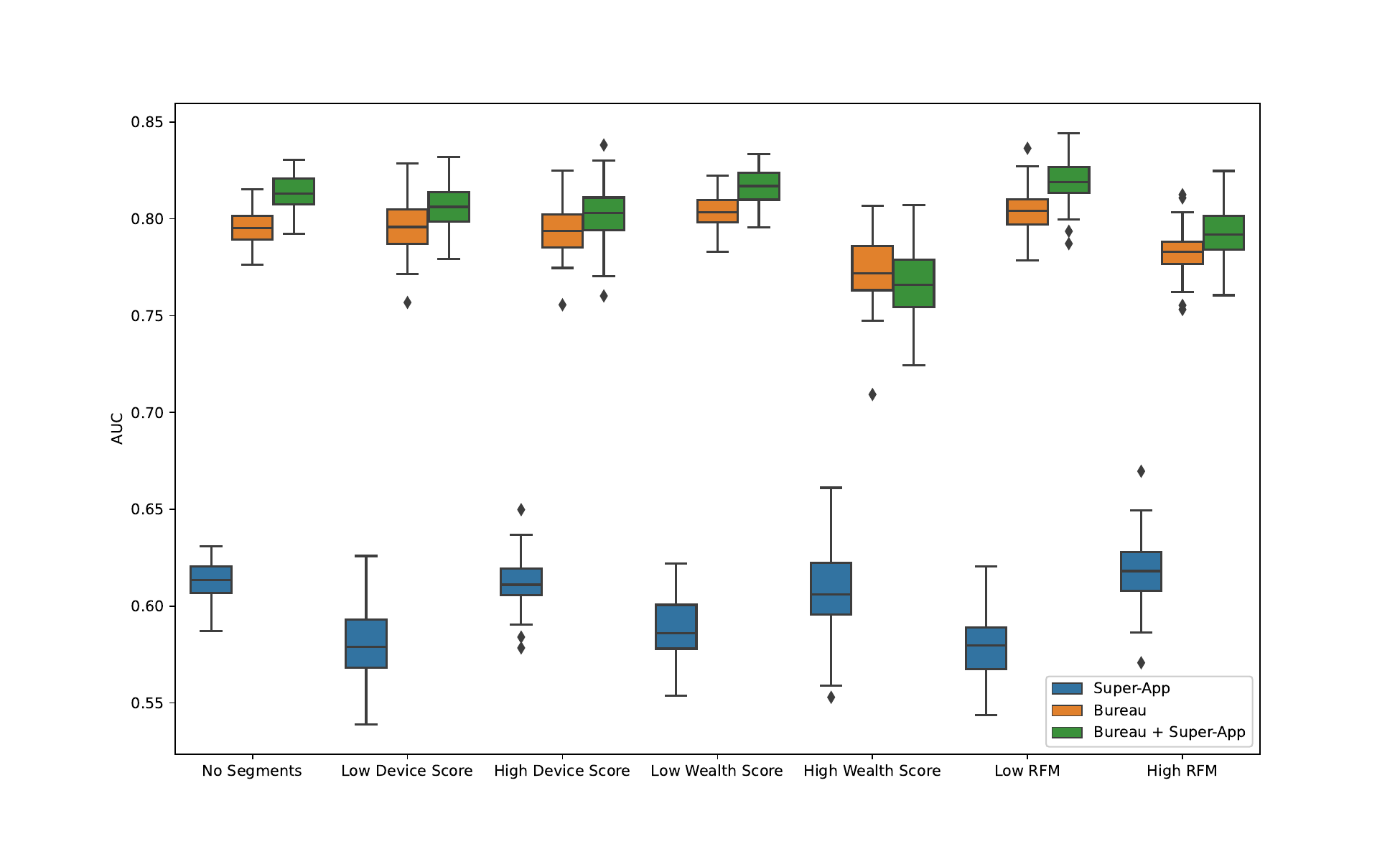}}
    \caption{Country B}
    \label{model_perf_auc_b}
\end{subfigure}
\caption{AUC performance by model.}
\end{figure}

The results obtained for both countries, according to the AUC metric, show the model that combines bureau and super-app variables achieves a higher average performance regardless of the segmentation as can be seen in Figures~\ref{model_perf_auc_a} and \ref{model_perf_auc_b}. For Country A, when all features are included the Device Score did appear to be a population characteristic that allowed for the better prediction of defaulters, given that for those with a high score a higher average performance was gained. However, for Country B this characteristic did not produce a substantial improvement in either the low or high scores. Nevertheless, the results associated with the Wealth Score and RFM segments suggest a significant difference in the performance of the users with a specific value for these characteristics for both countries. Although the model with only super-app features does not outperforms the bureau score in terms of prediction power, it manages to retrieve insights that the bureau score fails to capture. This explains why the combined model reaches the greatest predictive power.  Hence, future studies could evaluate whether a model built exclusively with super-app features could outperform traditional unbanked origination models, even if this only happen for special segments like High Device Score, High Wealth Score or High RFM, where it can be seen that predictive power stands out in comparison with other segments for country A.

\begin{figure}[tbp!]
\begin{subfigure}{\textwidth}
    \centerline{\includegraphics[scale = 0.40]{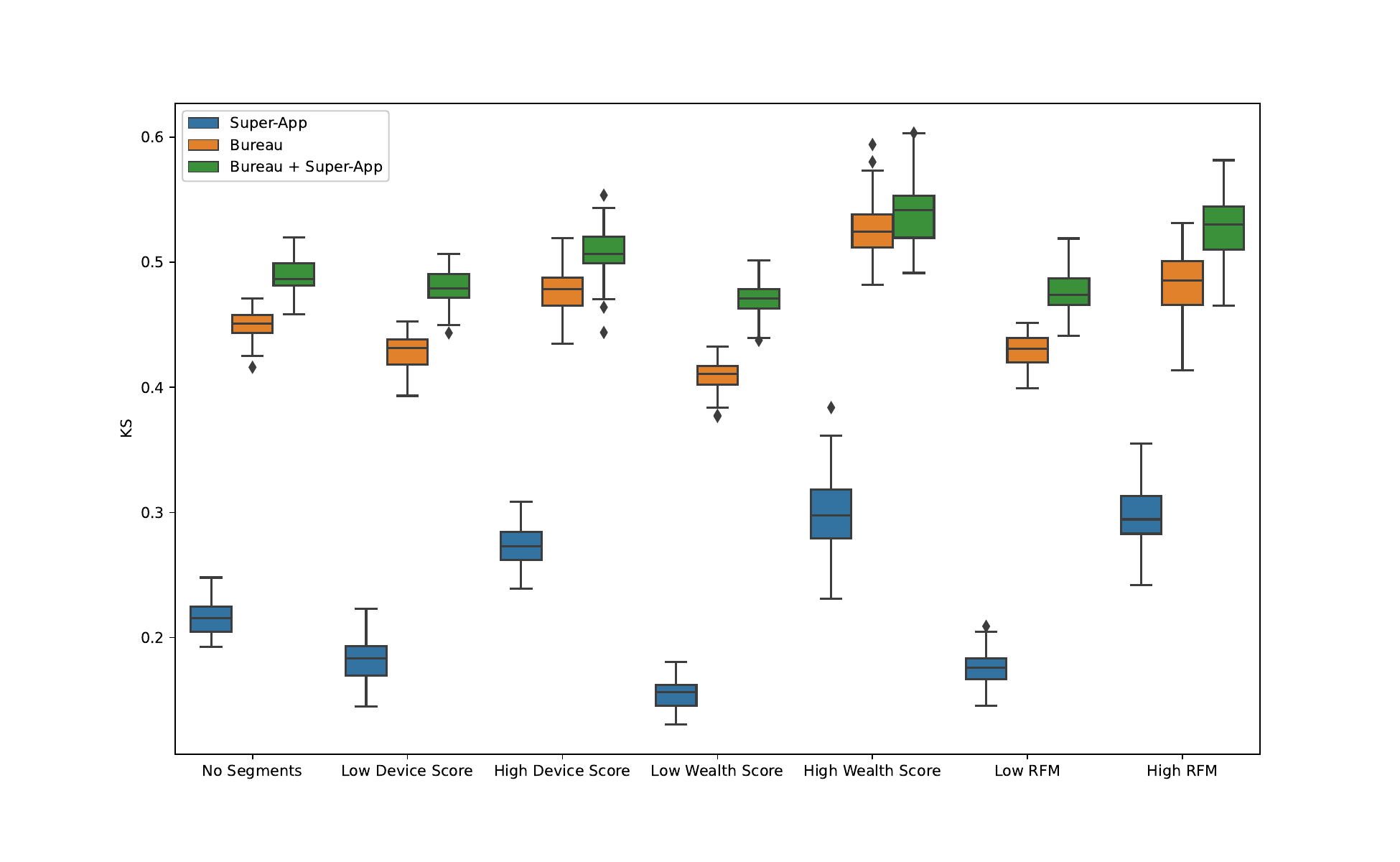}}
    \caption{Country B}
    \label{model_perf_ks_a}
\end{subfigure}

\begin{subfigure}{\textwidth}
    \centerline{\includegraphics[scale = 0.40]{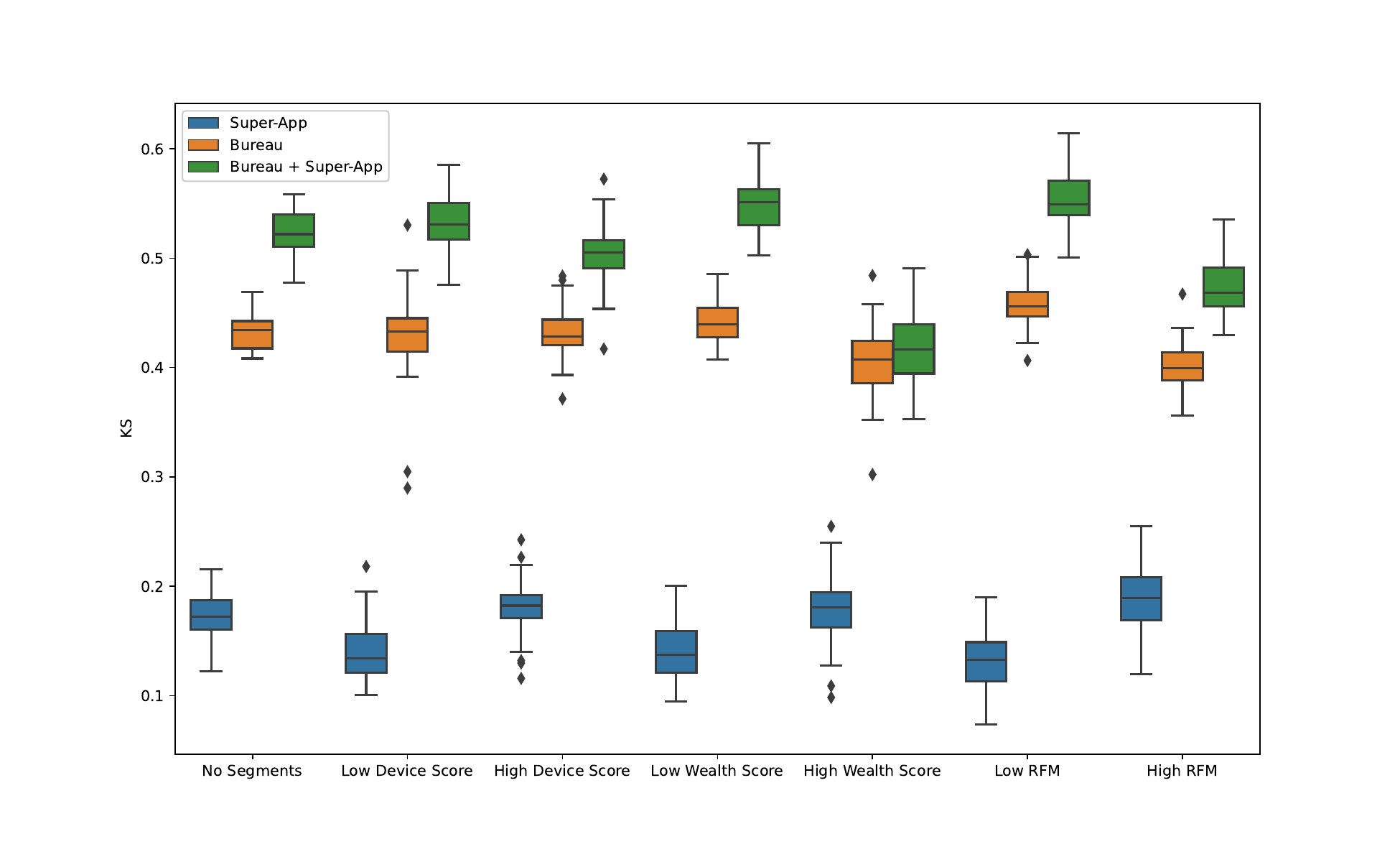}}
    \caption{Country B}
    \label{model_perf_ks_b}
\end{subfigure}
\caption{KS performance by model.}
\end{figure}

The results obtained with the KS metric were more pronounced as can be seen in Figures~\ref{model_perf_ks_a} and \ref{model_perf_ks_b}. For all the segments, and for both countries, the information from the super-app together with the bureau information makes the combined model outperform the bureau stand-alone model. Country B demonstrated a particularly high improvement with an average increase of approximately ten percentage points when adding super-app features to the bureau model. This implies that the super-app information has a significant impact upon the ability of the model to discriminate. The difference is most obvious in the Device Score (low) and Wealth Score (low), which hints at a much higher discrimination capacity for lower income segments. Something similar occurs in the RFM cluster, although with a lower effect. This could indeed occur as there should be a correlation between wealth and RFM, but it is muddled by the current engagement of the user with the app.

Although the box plots provide an approximation of the distribution of the performance measures, Mann Whitney non-parametric mean tests were conducted to more accurately determine which segmentation provided a significant improvement between the  super-app, bureau and combined models. Table \ref{Var_Mean_test} shows the non-zero p-values of the Mann Whitney test between the bureau and combined models. Thus, Table 6 implies that a significant difference is obtained by adding the super-app variables in all the models and segmentations for both countries and performance metrics, except for the increase in the K-S metric of the High Wealth model for Country B. We can also see that the models with the highest AUC performance were the High Wealth Score model for Country A and the Low RFM model for Country B, while for the K-S statistic it was the High and Low Wealth Scores, respectively.

\renewcommand{\baselinestretch}{1}
\renewcommand{\arraystretch}{1.1}
\begin{table}[th!]
\centering
\begin{tabular}{lcccc}
\toprule
& \multicolumn{2}{c}{Country A} & \multicolumn{2}{c}{Country B} \\
\midrule
Segmentation      & AUC       & KS       & AUC        & KS        \\
\hline
No Segments       & 1.762e-17 & 8.648e-18 & 2.078e-12 & 3.529e-18 \\
Low Device Score  & 4.771e-15 & 6.814e-18 & 9.204e-05 & 1.611e-17 \\
High Device Score & 2.248e-16 & 9.237e-11 & 1.280e-03 & 2.740e-16 \\
Low Wealth Score  & 4.295e-12 & 3.530e-18 & 8.375e-09 & 3.531e-18 \\
High Wealth Score & 1.137e-13 & 1.084e-02 & 3.110e-02 & \textbf{6.338e-02} \\
Low RFM           & 2.766e-15 & 1.659e-17 & 1.328e-07 & 3.980e-18 \\
High RFM          & 6.439e-14 & 2.090e-11 & 2.000e-04 & 2.102e-17 \\ 
\bottomrule
\end{tabular}
\caption{Mann Whitney test P-Value for performance metrics}
\label{Var_Mean_test}
\end{table}

\subsection{Financial model performance}
\begin{figure}[tbp!]
\begin{subfigure}{\textwidth}
    \centerline{\includegraphics[scale = 0.40]{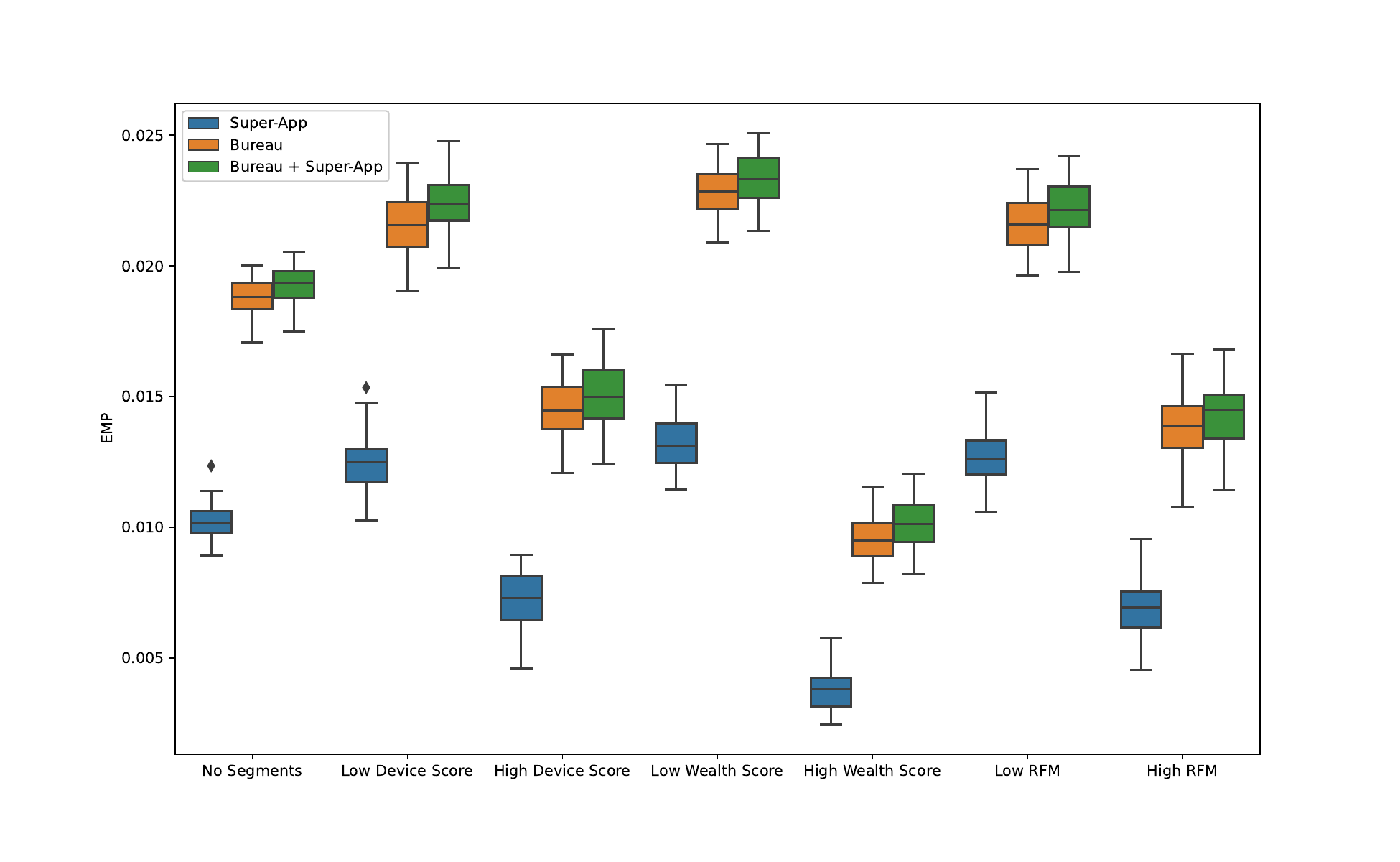}}
    \caption{Country A}
    \label{EMP_a}
\end{subfigure}

\begin{subfigure}{\textwidth}
    \centerline{\includegraphics[scale = 0.40]{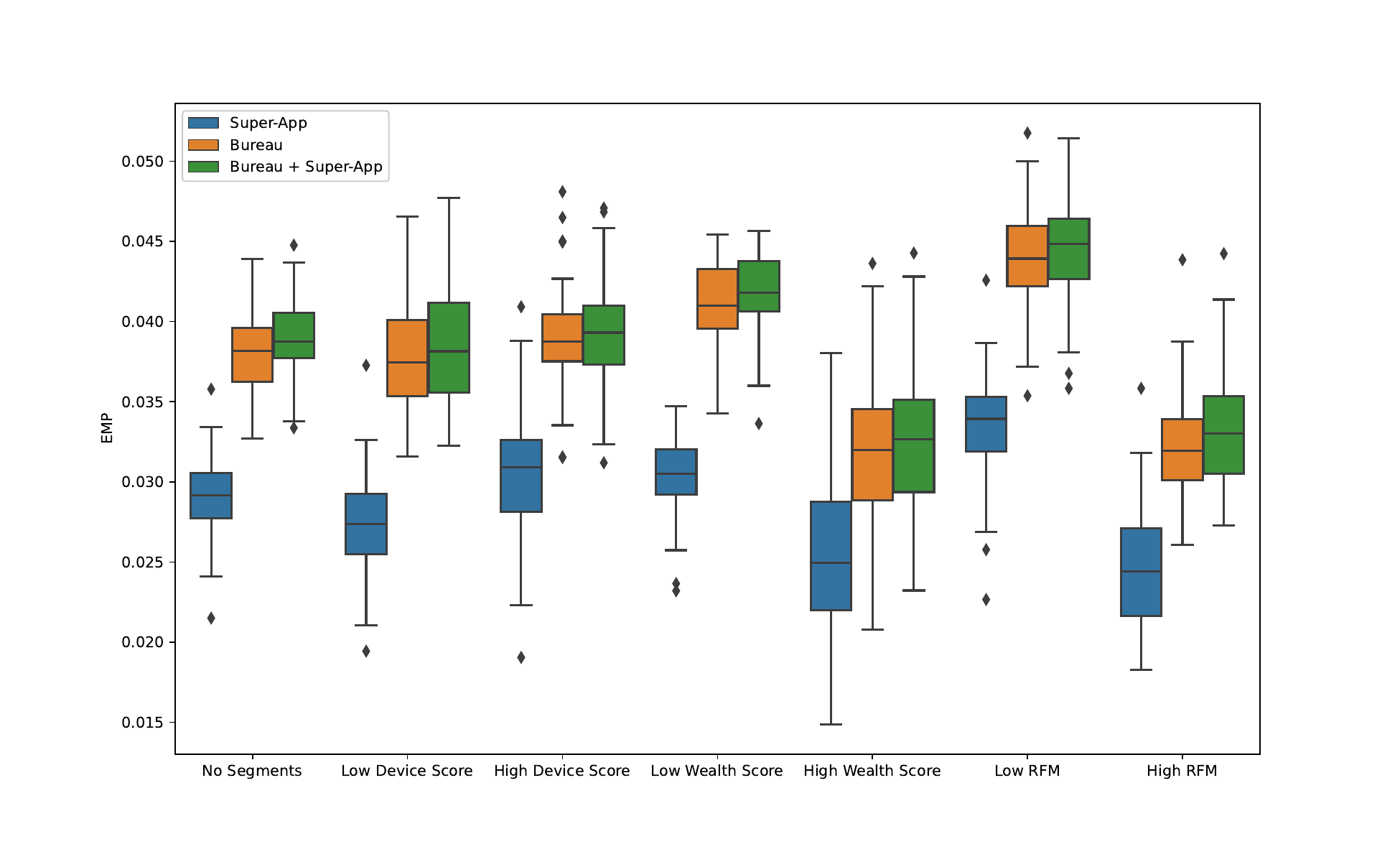}}
    \caption{Country B}
    \label{EMP_b}
\end{subfigure}
\caption{EMP performance by model.}
\end{figure}

In the previous section it was shown that the statistical performance of the model that combines super-app variables and bureau was significantly higher than the other models. However, the model that performs the best in terms of statistical measures does not necessarily performs the best in terms of costs and savings. The results obtained with the EMP approach are presented in Figures~\ref{EMP_a} and \ref{EMP_b}, for countries A and B respectively. As can be seen, regardless of the country or the segment, the model that considers only super-app features has the lowest average EMP, while the highest EMP is achieved by the combined model. Additionally, for both countries the largest EMP measures are found in the low-value segments, as would be expected given that the bureau often works better with high-value segments and super-app features enhance the financial performance of users for which the bureau does not have as much information.

Similar to the AUC and KS metrics, a Mann-Whitney non-parametric test is performed to identify significant increases in EMP. Table \ref{Mean_test_EMP} contains the p-values of the tests between the combined model and the model with only bureau features, the rest of p values are zero, showing a significant difference in means. Regarding the remaining tests, for country A there is a significant difference independent of the segment used, however for country B only the non-segment model manages to give a significant difference. This is related to the long-term view of the EMP measure and the availability of data for country B, but the rest of the tests strongly imply these differences will become significant as more data is available and finer-grained costs and benefits become known.

\begin{table}[tbh!]
\centering
\caption{Mann Whitney test test P-Value for EMP performance}
\begin{tabular}{lcccc}
\toprule
Segmentation      & Country A  &  Country B \\
\midrule
No Segments       & 0.00029  &  0.02413 \\
Low Device Score  & 0.00026  &  \textbf{0.14032} \\
High Device Score & 0.01883  &  \textbf{0.27778} \\
Low Wealth Score  & 0.00393  &  \textbf{0.13426} \\
High Wealth Score & 0.00137  &  \textbf{0.27778} \\
Low RFM           & 0.00491  &  \textbf{0.11054} \\
High RFM          & 0.03793  &  \textbf{0.06511} \\
\bottomrule
\end{tabular}
\label{Mean_test_EMP}
\end{table}

\begin{figure}[tbp!]
\begin{subfigure}{\textwidth}
    \centerline{\includegraphics[scale = 0.40]{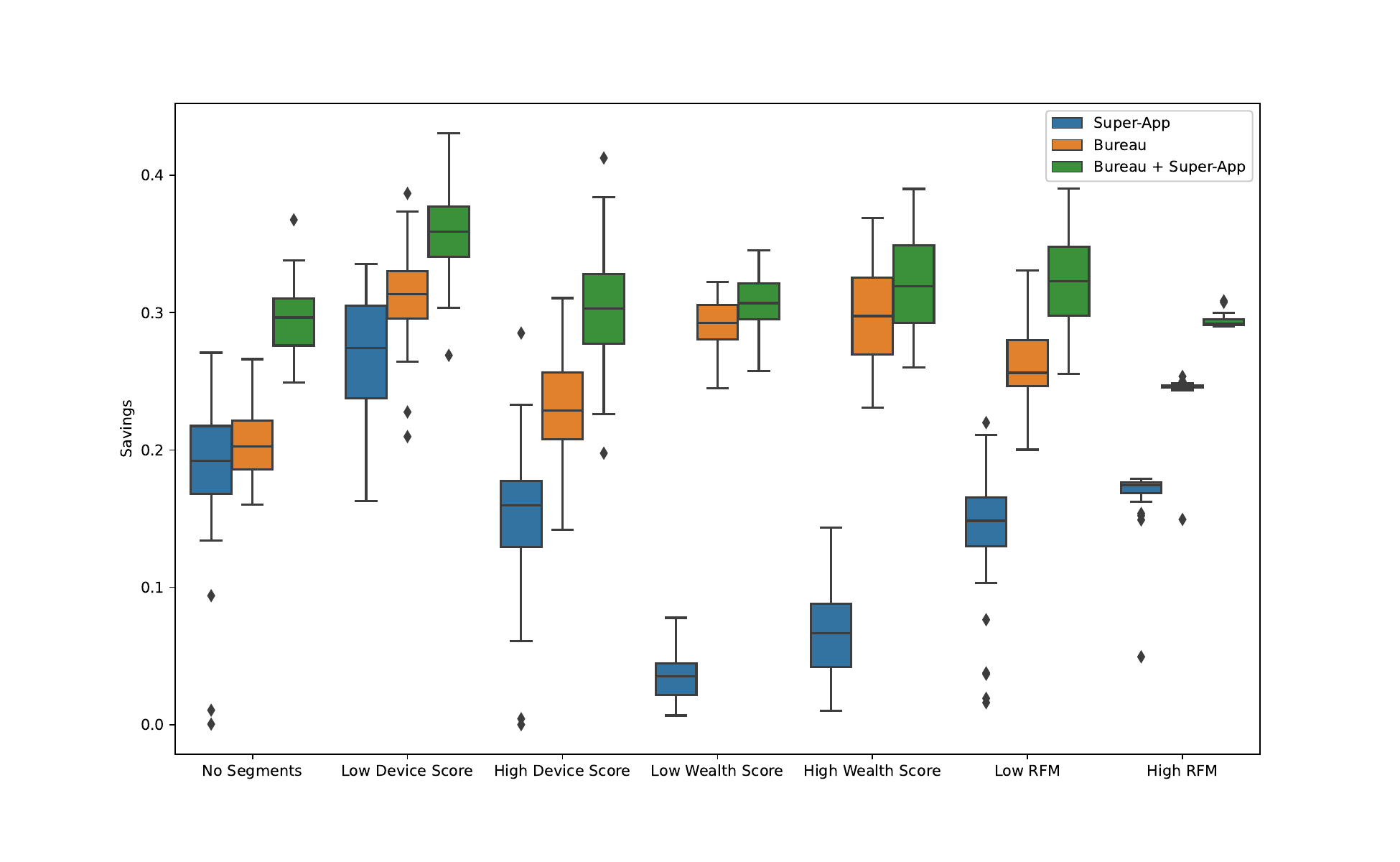}}
    \caption{Country A}
    \label{savings_a}
\end{subfigure}

\begin{subfigure}{\textwidth}
    \centerline{\includegraphics[scale = 0.40]{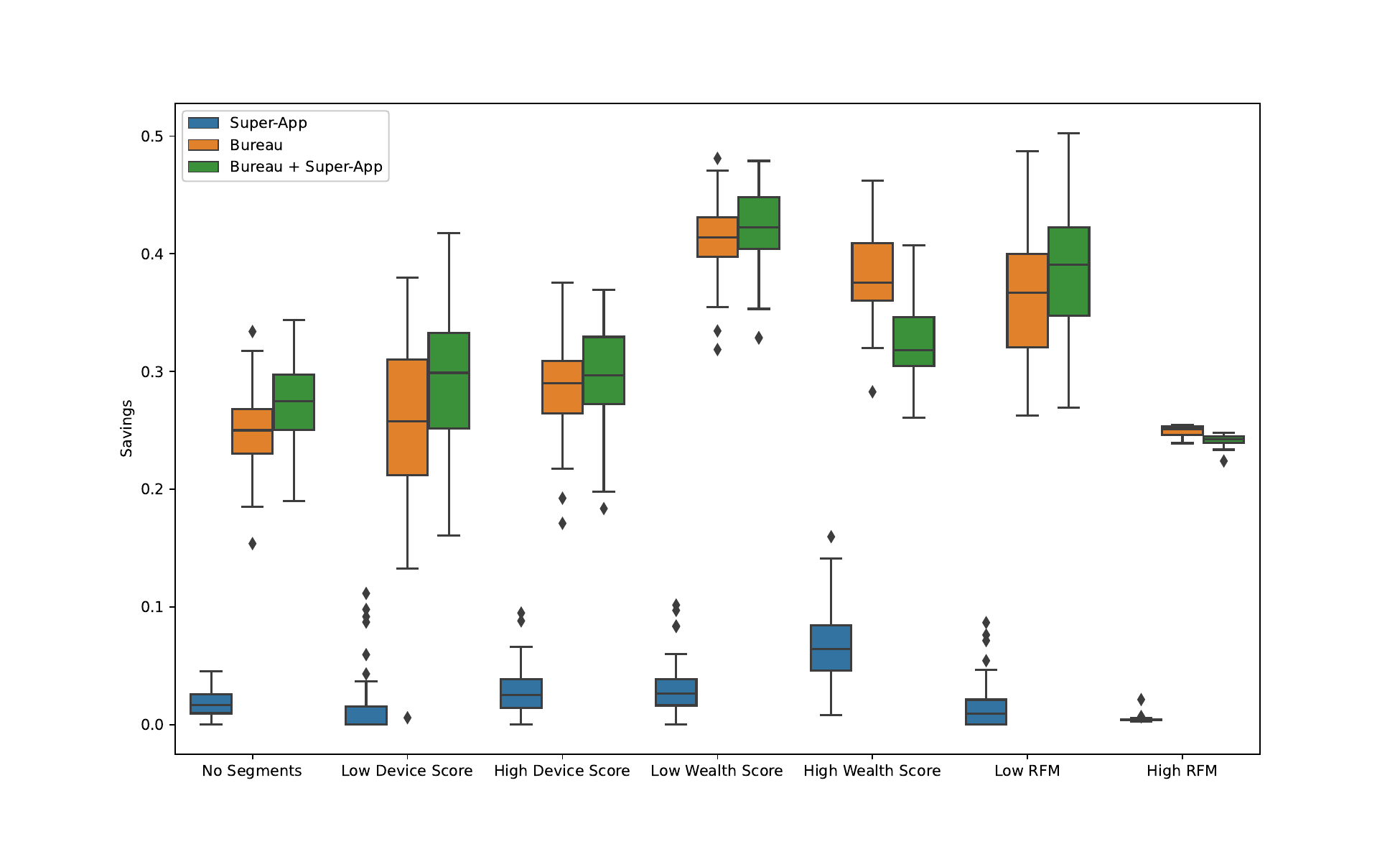}}
    \caption{Country B}
    \label{savings_b}
\end{subfigure}
\caption{Financial model performance.}
\end{figure}

Regarding \citet{CorreaBahnsen2014b} example-dependent cost-sensitive methodology's results, it was found that for both countries there was a higher average saving in all the models with combined variables, as well as considerable savings when segmenting the population. For Country A, as seen in Figure~\ref{savings_a}, the average saving when adding the super-app variables to bureau ones ranged from 20.5\%-31.3\% to 29.3\%-36.0\%, while for Country B, (Figure~\ref{savings_b}) there was a slight increase ranging from 25.0\%-41.0\% to 27.4\%-42.0\%.

The most accentuated financial differences for Country A were obtained for the (overall) no segment model and the models with High Device Score, Low RFM and High RFM segmentations; the respective average increases after adding the super-app variables were 9.0\%, 7.0\%, 5.9\% and 4.9\%. For country B, the largest average increases were provided by the Low Device Score, Low RFM and no segments models at values of 3.5\%, 2.2\% and 2.3\%, respectively. In addition, although for both countries these were not the models with the best average statistical performances, these results reveal not only the additional statistical value of the super-app variables, but also the financial benefit of considering the costs incurred during the default prediction process.

Overall, Figures \ref{savings_a} and \ref{savings_b} show a greater gain in terms of savings for Country A when adding alternative data. Table \ref{Mean_test_savings} shows the p-values of the Mann-Whitney non-parametric test only for the combined model as the p-value of the other test are zero. The test for the combined model implies that for Country B, the segments of High Device Score and Low Wealth Score did not significantly enhance savings. 

\begin{table}[tbh!]
\centering
\caption{Mann Whitney test test P-Value for financial performance}
\begin{tabular}{lcccc}
\toprule
Segmentation      & Country A  &  Country B \\
\midrule  
No Segments       & 6.828e-18  &  3.178e-04 \\
Low Device Score  & 6.158e-11  &  7.839e-03 \\
High Device Score & 5.458e-12  &  \textbf{8.451e-02} \\
Low Wealth Score  & 1.898e-04  &  \textbf{6.424e-02} \\
High Wealth Score & 1.791e-03  &  1.581e-10 \\
Low RFM           & 4.265e-13  &  1.883e-02 \\
High RFM          & 3.533e-18  &  1.979e-12 \\  
\bottomrule
\end{tabular}
\label{Mean_test_savings}
\end{table}

\subsection{Feature Importance}

Considering that a highly complex model was implemented, SHapley Additive exPlanations \citep[SHAP;][]{shap2017} was used for robust feature importance explanation, specifically TreeSHAP approach for tree-based models. Since this technique is based on game theory, specifically Shapley’s optimal values, it offers a unique way of consistently and precisely assigning importance to features and to acknowledge global and local interpretability.

\begin{figure}[tbp!]
\begin{subfigure}{\textwidth}
    \centerline{\includegraphics[scale = 0.35]{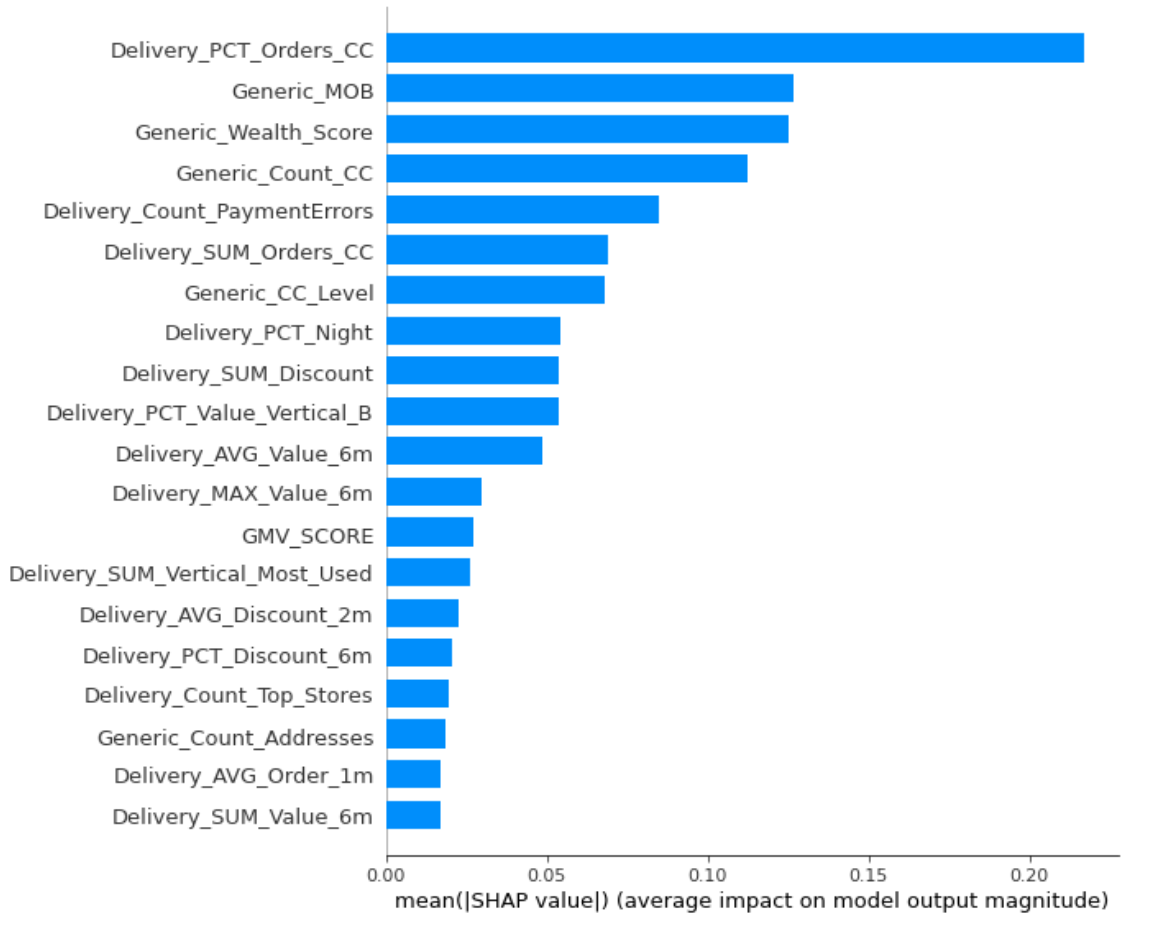}}
    \caption{Country A}
    \label{shapGR_a}
\end{subfigure}

\begin{subfigure}{\textwidth}
    \centerline{\includegraphics[scale = 0.35]{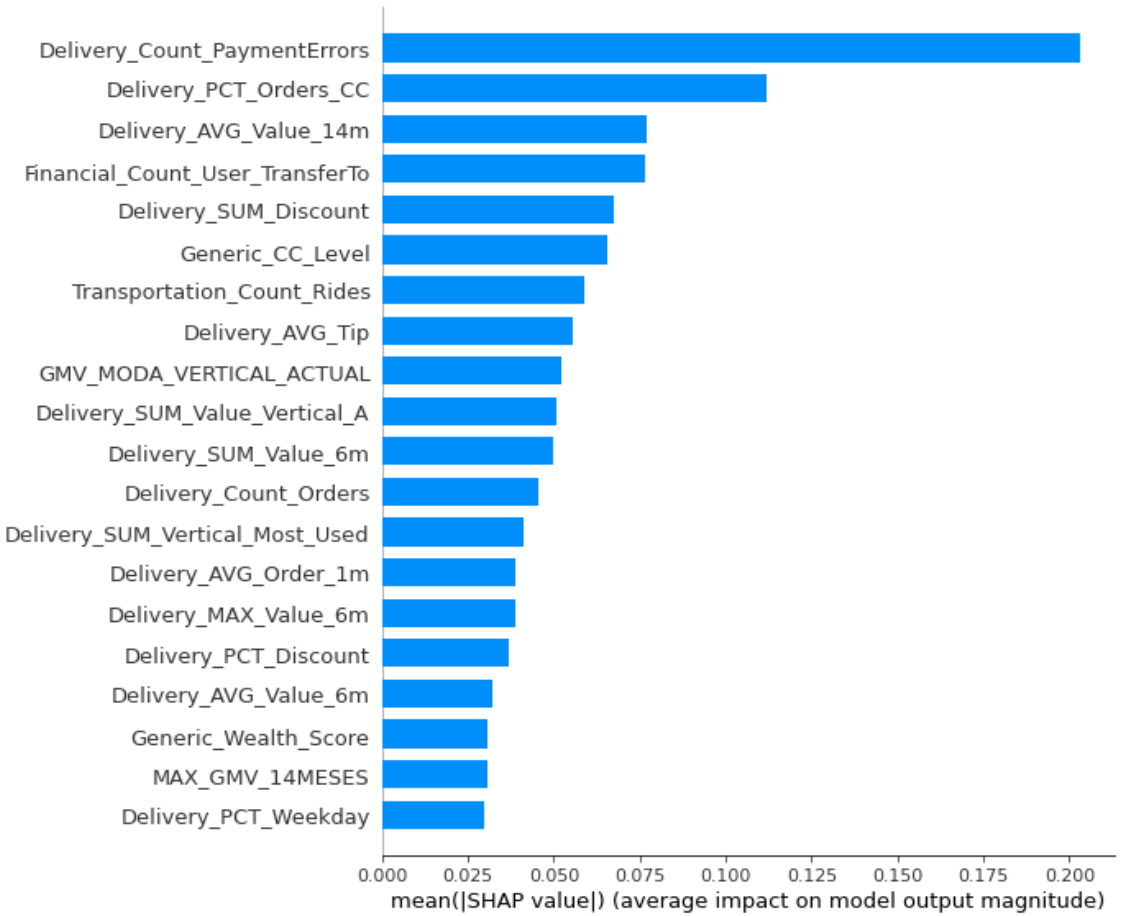}}
    \caption{Country B}
    \label{shapGR_b}
\end{subfigure}
\caption{Global feature importance with SHAP for both countries with super-app features only}
\end{figure}

\begin{figure}[tbp!]
\begin{subfigure}{\textwidth}
    \centerline{\includegraphics[scale = 0.40]{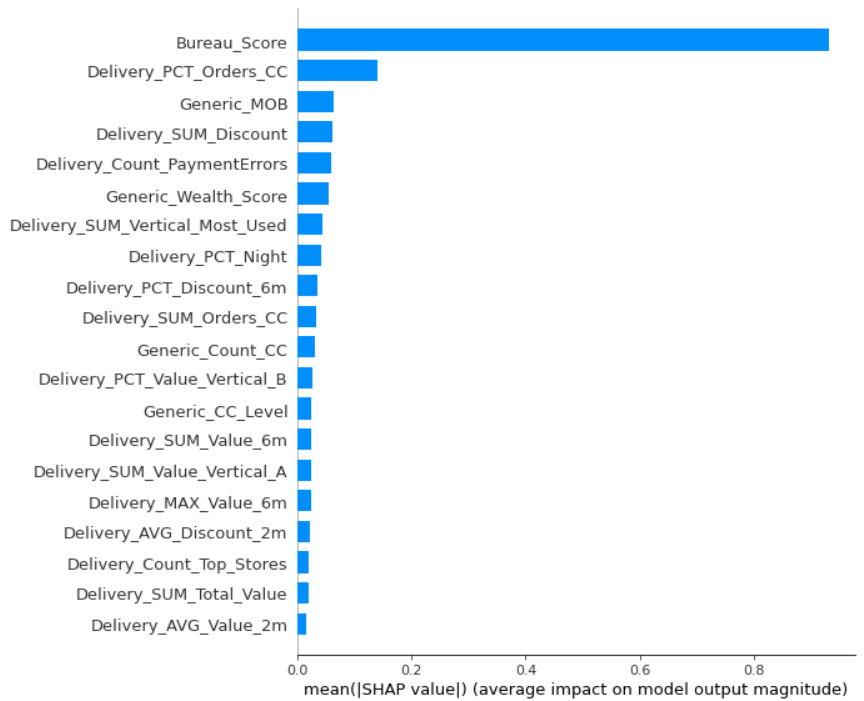}}
    \caption{Country A}
    \label{shapG_a}
\end{subfigure}

\begin{subfigure}{\textwidth}
    \centerline{\includegraphics[scale = 0.40]{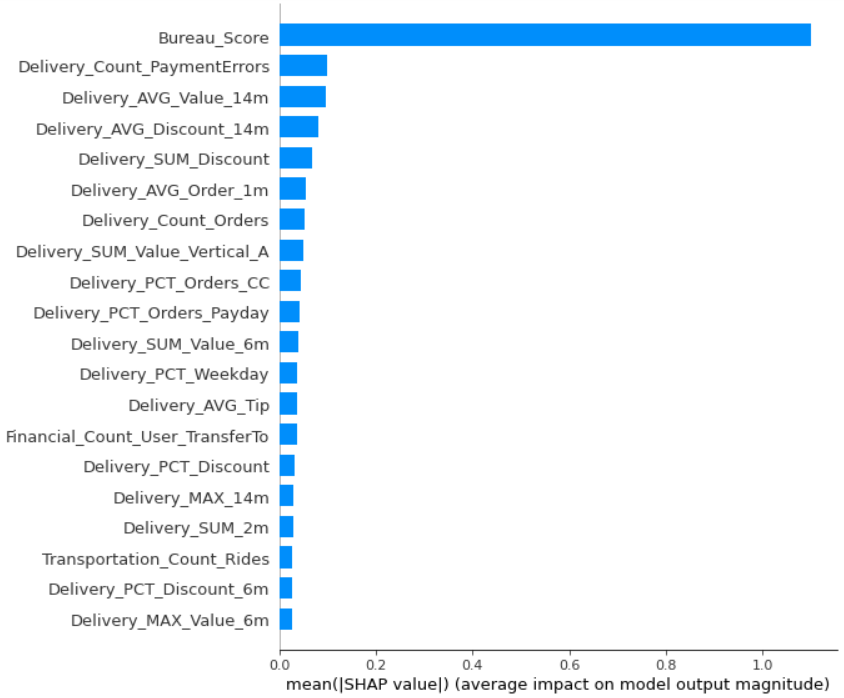}}
    \caption{Country B}
    \label{shapG_b}
\end{subfigure}
\caption{Global feature importance with SHAP for both countries with combined features}
\end{figure}

The global feature importance is obtained only for the  super-app variables model and the combined model, without considering segments. These results can be seen in Figures  ~\ref{shapGR_a} and \ref{shapG_a} for country A and in Figures~\ref{shapGR_b} and \ref{shapG_b}, for Country B. With the feature importance obtained with the TreeSHAP, is evident that for both countries when adding the bureau some of the super-app variable importance decrease. This variation is expected due to the fact that the bureau score manage to captures similar information that some of the super-app variables incorporates. Thus, variables such as Generic Wealth Score and Generic Count CC for country A and Financial Count User Transfer To and Generic CC level for country B lose predictive power as these capture financial information and economic potential of an user that the bureau captures in a more robust way (if this information is available). However, transactional variables such as the amount of payments with errors and orders paid with the super-app's own credit cards provide additional information to the one provided by bureau as they reflect behaviors not easily captured by financial data. 

Regarding the global importance of the combined features model, for both countries (Figures ~\ref{shapG_a} and \ref{shapG_b}), although the bureaus had the highest predictive power, super-app delivery and generic features achieved the most complementary performances to bureau information in the predictive power of creditworthiness. The latter super-app set reveals that although most sociodemographic (generic) features should be readily available to bureaus, the tenure and time of engagement are also extremely relevant in predicting default. These variables would relate mostly to each institution (as nothing suggests this variable would be exclusive to super-apps), hinting at the higher predictive power of in-house models over wider ranging bureau ones.

Regarding Country A the features associated with the payment method behavioral patterns were those that added more value to the accuracy of the default prediction. For Country B, all the categories of the delivery consumption patterns contributed in similar measures. Similarly, some financial and transportation features added value to the default prediction. Generic Month-on-Books (MOB) for Country A, and delivery payment errors, for both countries, were also relevant features.

\begin{figure}[tbp!]
\begin{subfigure}{\textwidth}
    \centerline{\includegraphics[scale = 0.45]{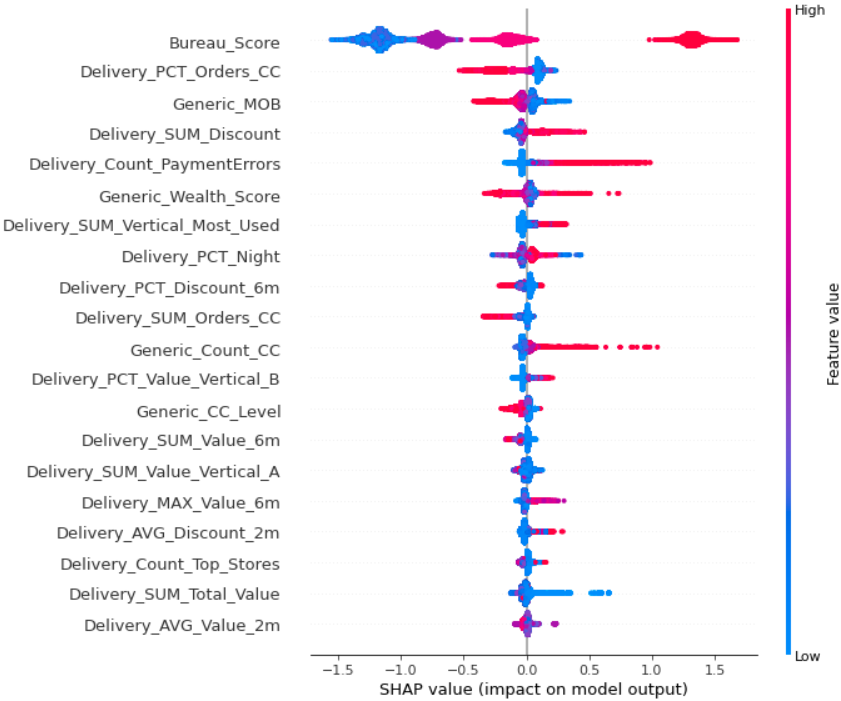}}
    \caption{Country A}
    \label{shap_a}
\end{subfigure}

\begin{subfigure}{\textwidth}
    \centerline{\includegraphics[scale = 0.45]{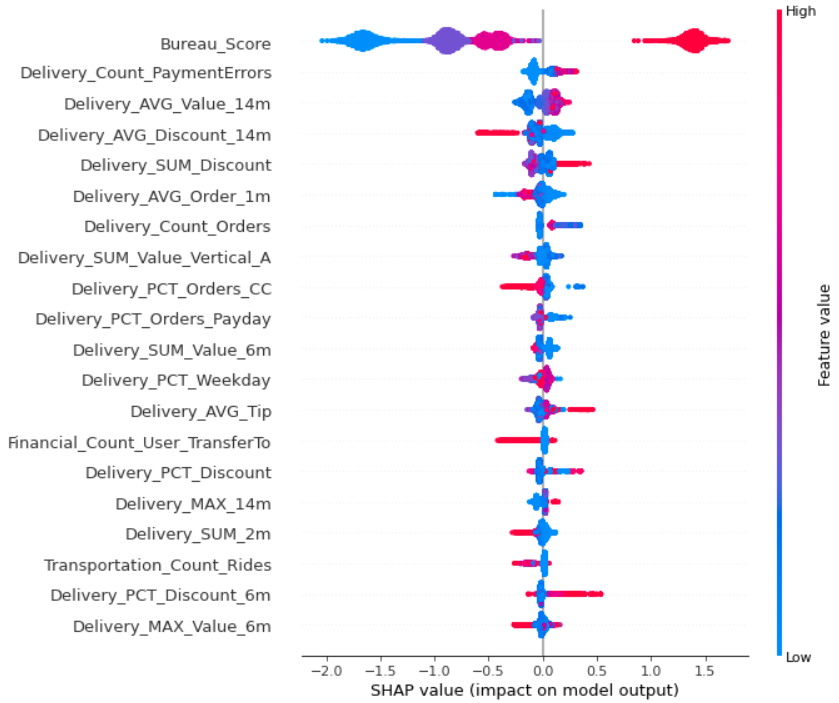}}
    \caption{Country B}
    \label{shap_b}
\end{subfigure}
\caption{Local interpretability with SHAP}
\end{figure}

Moreover, Figures~\ref{shap_a} and \ref{shap_b} present the impact of each feature on the model’s prediction. This local interpretability provided by the SHAP values allows us to understand how the value of a certain variable increases or decreases the probability of default, while also identifying non-linear relationships between all the variables. Hence, the local interpretability plots sorts features by global importance and uses SHAP values to measure the effect that each feature has on the model predictions.

Country A’s high MOB implies the greater likelihood of no default, as expected. These new sources of data consist of diverse behaviors, which pose the risk that the alternative data patterns could vary as times goes by. Accordingly, proper model validation and follow-up processes become even more relevant for these data sources.

For delivery payment error, the interpretation was the same for both countries as the number of errors increases so does the PD. This would mean there is an early warning of financial difficulties, as declines over time hint at a lower available disposable income.

Most of other variables revealed interesting behaviors that are not captured by traditional bureau variables. In particular those variables that involved credit card use or utilization (those ending in CC),  were among the most relevant sets. These can be interpreted as bancarization engagement indexes: users with a higher level of engagement are in general better customers, however, if the amount is too high then, \emph{ceteris paribus}, the borrower has an increased default rate.

Finally, the last interesting result arising from these variable importance plots derives from the magnitude and range of the SHAP values. Considering that the higher the SHAP value, the more impact a variable has on the output, it is clear that extremely low and extremely high bureau scores (shown in blue and red in the Bureau Score row of the plot respectively) are the most impactful variables. This is well-known: it is much easier to predict a very good or a very bad borrower than an average one. It is in this segment that the super-app variables really shine. Bureau scores do not have small difference ranges for the defaulters, that is, a range of scores that allows the discrimination of slightly bad borrowers on a sliding scale. However, variables such as Payment Errors and Total Amount do exactly that. This allows the fintech to take calculated risks by accepting slightly riskier borrowers for a temporary income boost (such as a boosted growth phase common in technology companies), with which traditional banks would not be able to compete.

\subsection{Regulatory Implications}\label{sec:Regulatory}

There are many interesting lessons for regulators arising from this work. First, the statistical and financial gains resulting from both segmented and non-segmented models demonstrate that alternative data have a place in the lending sphere and, furthermore, that lenders have a financial incentive to use these variables within their models. In recent reports, the Basel Committee on Banking Supervision detected this trend \citep{BCBSFintech2018} and has suggested that regulators should treat these fintech companies in a similar manner to banks from a regulatory perspective. Our results suggest they should progress a step further by encouraging this information to become mainstream, which we propose could lead to higher bancarization and more general access to financing rates.

However, the counterpoint to these gains is that they must come with a clear mandate concerning the interpretability of the variables. Clear arguments have to be presented on exactly what the variables are illustrating and how they relate to financial behaviors. For example, this is the case with a variable such as tipping, which had a slightly negative effect in our model. Experiments have shown that psychometric variables, to which tipping behavior is related, have an impact upon the creditworthiness of borrowers \citep{irani2017}, to which the tipping behaviour is related to, but what exactly what is this is showing has to be clearly explained by the lender in order to be approved for use in scoring models. In this exploratory study, the variable indicated that high tippers have a significantly higher default rate. Allowing the use of such a variable would provide an incentive for users to stop tipping altogether (although small tips had a significant but lower impact upon lower default rates in contrast), which is an undesirable consequence. Thus, the regulator should intervene when case variables such as this are proposed and potentially forbid their use or control how they are used. In \citet{Oskarsdottir2019}, a suggestion for the use of such variables involved using only the positive part, that is, considering only the segment of the variable that is positive and fixing a neutral score to those that are not. Doing so rewards positive behaviors, while eliminating the impact of more dubious ones in terms of why the phenomenon occurs. However, further research needs to be conducted to understand the underlying reasoning of these results to arrive at a final recommendation regarding these variables.

\section{Conclusions}

In this paper, we have tested an alternative dataset arising from a super-app and researched its effectiveness and implications with regard to developing credit risk models. Four research questions were proposed and our study clearly answered each of them.

First, there is clear predictive value both in financial and in statistical terms in using these variables, which answers the first two research questions. The gains were significant across all the studied segments, and these gains were consistent across both the countries in which we tested these variables. Clearly, a financial and statistical incentive exists for lenders to include app-based information in their decision support systems. These conclusions hold, for most models and datasets, both in terms of the EMP and the EDCS measure.

Typically, alternative data variables are strong indicators of both  the willingness and capacity to repay a loan \citep{Bravo_2015}. However, the types and varieties of these indicators (almost 20 new variables had significant effects upon the estimation) result in significant gains in both financial and statistical terms. This paper demonstrates that there is a strong financial incentive for financial institutions to use these variables for prediction models. As the financial incentive is high, this means for a while fintech companies will have an advantage over traditional institutions unless these institutions also begin embracing these new sources of information.

Regarding the third question, in terms of the patterns we observed, the super-app variables show that engagement with financial products provide the strongest signals in terms of the default rate prediction. These patterns did not appear to be readily included in bureau scores as they do not collect signs of debt but of transactionality, hence, there is an opportunity for them to be incorporated in the mainstream. In the meantime, those institutions with access to these variables have a competitive advantage with regard to designing better decision support systems for this purpose.

Finally, we foresee regulators will have to balance allowing this data to be used with effective supervision over what patterns these variables actually reflect. Given the financial incentives that arise from the use of these variables, it will be necessary to take measures to safeguard a fair and transparent app-based lending system. Nonetheless, our results suggest that these apps are useful contributions to financial inclusion, therefore, regulatory efforts should also proceed in this direction.

\section*{Acknowledgments}
The last author acknowledges this research was undertaken, in part, thanks to funding from the Canada Research Chairs program.


\begin{thebibliography}{}

\bibitem [\protect \citeauthoryear {%
Aitken%
}{%
Aitken%
}{%
{\protect \APACyear {2017}}%
}]{%
Aitken2017}
\APACinsertmetastar {%
Aitken2017}%
\begin{APACrefauthors}%
Aitken, R.%
\end{APACrefauthors}%
\unskip\
\newblock
\APACrefYearMonthDay{2017}{}{}.
\newblock
{\BBOQ}\APACrefatitle {‘All data is credit data’: Constituting the
  unbanked} {‘all data is credit data’: Constituting the unbanked}.{\BBCQ}
\newblock
\APACjournalVolNumPages{Competition \& Change}{21}{4}{274-–300}.
\PrintBackRefs{\CurrentBib}

\bibitem [\protect \citeauthoryear {%
Arr{\'a}iz%
, Bruhn%
, Ortega%
\BCBL {}\ \BBA {} Stucchi%
}{%
Arr{\'a}iz%
\ \protect \BOthers {.}}{%
{\protect \APACyear {2017}}%
}]{%
irani2017}
\APACinsertmetastar {%
irani2017}%
\begin{APACrefauthors}%
Arr{\'a}iz, I.%
, Bruhn, M.%
, Ortega, C\BPBI R.%
\BCBL {}\ \BBA {} Stucchi, R.%
\end{APACrefauthors}%
\unskip\
\newblock
\APACrefYearMonthDay{2017}{{\APACmonth{12}}}{}.
\newblock
\APACrefbtitle {Are {{Psychometric Tools}} a {{Viable Screening Method}} for
  {{Small}} and {{Medium}}-{{Size Enterprise Lending}}? {{Evidence}} from
  {{Peru}}} {Are {{Psychometric Tools}} a {{Viable Screening Method}} for
  {{Small}} and {{Medium}}-{{Size Enterprise Lending}}? {{Evidence}} from
  {{Peru}}}\ \APACbVolEdTR{}{\BTR{}\ \BNUM\ 8276}.
\newblock
\APACaddressInstitution{}{{The World Bank}}.
\PrintBackRefs{\CurrentBib}

\bibitem [\protect \citeauthoryear {%
{Asian Insights Office }%
}{%
{Asian Insights Office }%
}{%
{\protect \APACyear {2019}}%
}]{%
DBS2019}
\APACinsertmetastar {%
DBS2019}%
\begin{APACrefauthors}%
{Asian Insights Office }.%
\end{APACrefauthors}%
\unskip\
\newblock
\APACrefYearMonthDay{2019}{{\APACmonth{09}}}{}.
\newblock
\APACrefbtitle {Super Apps in Financial Services: Business Models and
  Opportunities} {Super apps in financial services: Business models and
  opportunities}\ \APACbVolEdTR{}{\BTR{}\ \BNUM\ sector briefing 81}.
\newblock
\APACaddressInstitution{}{{DBS Group Research}}.
\PrintBackRefs{\CurrentBib}

\bibitem [\protect \citeauthoryear {%
Baesens%
, Roesch%
\BCBL {}\ \BBA {} Scheule%
}{%
Baesens%
\ \protect \BOthers {.}}{%
{\protect \APACyear {2016}}%
}]{%
baesens2016}
\APACinsertmetastar {%
baesens2016}%
\begin{APACrefauthors}%
Baesens, B.%
, Roesch, D.%
\BCBL {}\ \BBA {} Scheule, H.%
\end{APACrefauthors}%
\unskip\
\newblock
\APACrefYear{2016}.
\newblock
\APACrefbtitle {Credit Risk Analytics: Measurement Techniques, Applications,
  and Examples in {{SAS}}} {Credit risk analytics: Measurement techniques,
  applications, and examples in {{SAS}}}.
\newblock
\APACaddressPublisher{Hoboken, New Jersey}{{Wiley}}.
\PrintBackRefs{\CurrentBib}

\bibitem [\protect \citeauthoryear {%
{Basel Committee on Banking Supervision}%
}{%
{Basel Committee on Banking Supervision}%
}{%
{\protect \APACyear {2018}}%
}]{%
BCBSFintech2018}
\APACinsertmetastar {%
BCBSFintech2018}%
\begin{APACrefauthors}%
{Basel Committee on Banking Supervision}.%
\end{APACrefauthors}%
\unskip\
\newblock
\APACrefYearMonthDay{2018}{}{}.
\newblock
\APACrefbtitle {Sound {{Practices}}: Implications of Fintech Developments for
  Banks and Bank Supervisors} {Sound {{Practices}}: Implications of fintech
  developments for banks and bank supervisors}\ \APACbVolEdTR{}{\BTR{}}.
\newblock
\APACaddressInstitution{}{{Bank for International Settlements}}.
\PrintBackRefs{\CurrentBib}

\bibitem [\protect \citeauthoryear {%
Berg%
, Burg%
, Gombovi{\'c}%
\BCBL {}\ \BBA {} Puri%
}{%
Berg%
\ \protect \BOthers {.}}{%
{\protect \APACyear {2019}}%
}]{%
Berg2019}
\APACinsertmetastar {%
Berg2019}%
\begin{APACrefauthors}%
Berg, T.%
, Burg, V.%
, Gombovi{\'c}, A.%
\BCBL {}\ \BBA {} Puri, M.%
\end{APACrefauthors}%
\unskip\
\newblock
\APACrefYearMonthDay{2019}{}{}.
\newblock
{\BBOQ}\APACrefatitle {On the {{Rise}} of {{FinTechs}}: {{Credit Scoring Using
  Digital Footprints}}} {On the {{Rise}} of {{FinTechs}}: {{Credit Scoring
  Using Digital Footprints}}}.{\BBCQ}
\newblock
\APACjournalVolNumPages{The Review of Financial Studies}{Accepted for
  publication}{}{}.
\PrintBackRefs{\CurrentBib}

\bibitem [\protect \citeauthoryear {%
Bravo%
, Maldonado%
\BCBL {}\ \BBA {} Weber%
}{%
Bravo%
\ \protect \BOthers {.}}{%
{\protect \APACyear {2013}}%
}]{%
Bravo_2013}
\APACinsertmetastar {%
Bravo_2013}%
\begin{APACrefauthors}%
Bravo, C.%
, Maldonado, S.%
\BCBL {}\ \BBA {} Weber, R.%
\end{APACrefauthors}%
\unskip\
\newblock
\APACrefYearMonthDay{2013}{}{}.
\newblock
{\BBOQ}\APACrefatitle {Granting and Managing Loans for Micro-Entrepreneurs:
  {{New}} Developments and Practical Experiences} {Granting and managing loans
  for micro-entrepreneurs: {{New}} developments and practical
  experiences}.{\BBCQ}
\newblock
\APACjournalVolNumPages{European Journal of Operational
  Research}{227}{}{358--366}.
\PrintBackRefs{\CurrentBib}

\bibitem [\protect \citeauthoryear {%
Bravo%
, Thomas%
\BCBL {}\ \BBA {} Weber%
}{%
Bravo%
\ \protect \BOthers {.}}{%
{\protect \APACyear {2015}}%
}]{%
Bravo_2015}
\APACinsertmetastar {%
Bravo_2015}%
\begin{APACrefauthors}%
Bravo, C.%
, Thomas, C\BPBI L.%
\BCBL {}\ \BBA {} Weber, R.%
\end{APACrefauthors}%
\unskip\
\newblock
\APACrefYearMonthDay{2015}{}{}.
\newblock
{\BBOQ}\APACrefatitle {Improving Credit Scoring by Differentiating Defaulter
  Behaviour} {Improving credit scoring by differentiating defaulter
  behaviour}.{\BBCQ}
\newblock
\APACjournalVolNumPages{Journal of the Operational Research
  Society}{66}{5}{771--781}.
\PrintBackRefs{\CurrentBib}

\bibitem [\protect \citeauthoryear {%
Carroll%
\ \BBA {} Rehmani%
}{%
Carroll%
\ \BBA {} Rehmani%
}{%
{\protect \APACyear {2017}}%
}]{%
Carroll2017}
\APACinsertmetastar {%
Carroll2017}%
\begin{APACrefauthors}%
Carroll, P.%
\BCBT {}\ \BBA {} Rehmani, S.%
\end{APACrefauthors}%
\unskip\
\newblock
\APACrefYearMonthDay{2017}{}{}.
\newblock
\APACrefbtitle {Altenative Data and the Unbanked} {Altenative data and the
  unbanked}\ \APACbVolEdTR{}{\BTR{}}.
\newblock
\APACaddressInstitution{}{Oliver Wyman Financial Services}.
\PrintBackRefs{\CurrentBib}

\bibitem [\protect \citeauthoryear {%
{Correa Bahnsen}%
, Aouada%
\BCBL {}\ \BBA {} Ottersten%
}{%
{Correa Bahnsen}%
\ \protect \BOthers {.}}{%
{\protect \APACyear {2014}}%
}]{%
CorreaBahnsen2014b}
\APACinsertmetastar {%
CorreaBahnsen2014b}%
\begin{APACrefauthors}%
{Correa Bahnsen}, A.%
, Aouada, D.%
\BCBL {}\ \BBA {} Ottersten, B.%
\end{APACrefauthors}%
\unskip\
\newblock
\APACrefYearMonthDay{2014}{}{}.
\newblock
{\BBOQ}\APACrefatitle {{Example-Dependent Cost-Sensitive Logistic Regression
  for Credit Scoring}} {{Example-Dependent Cost-Sensitive Logistic Regression
  for Credit Scoring}}.{\BBCQ}
\newblock
\BIn{} \APACrefbtitle {2014 13th International Conference on Machine Learning
  and Applications} {2014 13th international conference on machine learning and
  applications}\ (\BPGS\ 263--269).
\newblock
\APACaddressPublisher{Detroit, USA}{IEEE}.
\PrintBackRefs{\CurrentBib}

\bibitem [\protect \citeauthoryear {%
{Correa Bahnsen}%
, Aouada%
\BCBL {}\ \BBA {} Ottersten%
}{%
{Correa Bahnsen}%
\ \protect \BOthers {.}}{%
{\protect \APACyear {2015}}%
}]{%
CorreaBahnsen2015}
\APACinsertmetastar {%
CorreaBahnsen2015}%
\begin{APACrefauthors}%
{Correa Bahnsen}, A.%
, Aouada, D.%
\BCBL {}\ \BBA {} Ottersten, B.%
\end{APACrefauthors}%
\unskip\
\newblock
\APACrefYearMonthDay{2015}{}{}.
\newblock
{\BBOQ}\APACrefatitle {{Example-Dependent Cost-Sensitive Decision Trees}}
  {{Example-Dependent Cost-Sensitive Decision Trees}}.{\BBCQ}
\newblock
\APACjournalVolNumPages{Expert Systems with Applications}{42}{19}{6609--6619}.
\PrintBackRefs{\CurrentBib}

\bibitem [\protect \citeauthoryear {%
{Dabla-Norris}%
\ \protect \BOthers {.}}{%
{Dabla-Norris}%
\ \protect \BOthers {.}}{%
{\protect \APACyear {2015}}%
}]{%
dabla-norris2015}
\APACinsertmetastar {%
dabla-norris2015}%
\begin{APACrefauthors}%
{Dabla-Norris}, E.%
, Deng, Y.%
, Ivanova, A.%
, Karpowicz, I.%
, Unsal, F.%
, Van~Leemput, E.%
\BCBL {}\ \BBA {} Wong, J.%
\end{APACrefauthors}%
\unskip\
\newblock
\APACrefYearMonthDay{2015}{}{}.
\newblock
\APACrefbtitle {Financial {{Inclusion}}: {{Zooming}} in on {{Latin America}}}
  {Financial {{Inclusion}}: {{Zooming}} in on {{Latin America}}}\
  \APACbVolEdTR{}{\BTR{}\ \BNUM\ WP/15/206}.
\newblock
\APACaddressInstitution{}{{International Monetary Fund}}.
\PrintBackRefs{\CurrentBib}

\bibitem [\protect \citeauthoryear {%
Demirgüç-Kunt%
, Klapper%
, Singer%
, Ansar%
\BCBL {}\ \BBA {} Hess%
}{%
Demirgüç-Kunt%
\ \protect \BOthers {.}}{%
{\protect \APACyear {2018}}%
}]{%
Asli2018}
\APACinsertmetastar {%
Asli2018}%
\begin{APACrefauthors}%
Demirgüç-Kunt, A.%
, Klapper, L.%
, Singer, D.%
, Ansar, S.%
\BCBL {}\ \BBA {} Hess, J.%
\end{APACrefauthors}%
\unskip\
\newblock
\APACrefYearMonthDay{2018}{}{}.
\newblock
\APACrefbtitle {The Global Findex Database 2017 : Measuring Financial Inclusion
  and the Fintech Revolution} {The global findex database 2017 : Measuring
  financial inclusion and the fintech revolution}\ \APACbVolEdTR{}{\BTR{}\
  \BNUM\ 126033}.
\newblock
\APACaddressInstitution{}{The World Bank}.
\newblock
\begin{APACrefDOI} \doi{10.1596/978-1-4648-1259-0} \end{APACrefDOI}
\PrintBackRefs{\CurrentBib}

\bibitem [\protect \citeauthoryear {%
Fader%
, Hardie%
\BCBL {}\ \BBA {} Lee%
}{%
Fader%
\ \protect \BOthers {.}}{%
{\protect \APACyear {2005}}%
}]{%
fader2005}
\APACinsertmetastar {%
fader2005}%
\begin{APACrefauthors}%
Fader, P\BPBI S.%
, Hardie, B\BPBI G\BPBI S.%
\BCBL {}\ \BBA {} Lee, K\BPBI L.%
\end{APACrefauthors}%
\unskip\
\newblock
\APACrefYearMonthDay{2005}{}{}.
\newblock
{\BBOQ}\APACrefatitle {{{RFM}} and {{CLV}}: {{Using Iso}}-{{Value Curves}} for
  {{Customer Base Analysis}}} {{{RFM}} and {{CLV}}: {{Using Iso}}-{{Value
  Curves}} for {{Customer Base Analysis}}}.{\BBCQ}
\newblock
\APACjournalVolNumPages{Journal of Marketing Research}{42}{4}{415--430}.
\PrintBackRefs{\CurrentBib}

\bibitem [\protect \citeauthoryear {%
Gool%
, Verbeke%
, Sercu%
\BCBL {}\ \BBA {} Baesens%
}{%
Gool%
\ \protect \BOthers {.}}{%
{\protect \APACyear {2012}}%
}]{%
gool2012}
\APACinsertmetastar {%
gool2012}%
\begin{APACrefauthors}%
Gool, J\BPBI V.%
, Verbeke, W.%
, Sercu, P.%
\BCBL {}\ \BBA {} Baesens, B.%
\end{APACrefauthors}%
\unskip\
\newblock
\APACrefYearMonthDay{2012}{}{}.
\newblock
{\BBOQ}\APACrefatitle {Credit Scoring for Microfinance: Is It Worth It?}
  {Credit scoring for microfinance: Is it worth it?}{\BBCQ}
\newblock
\APACjournalVolNumPages{International Journal of Finance \&
  Economics}{17}{2}{103--123}.
\PrintBackRefs{\CurrentBib}

\bibitem [\protect \citeauthoryear {%
Guo%
\ \protect \BOthers {.}}{%
Guo%
\ \protect \BOthers {.}}{%
{\protect \APACyear {2016}}%
}]{%
guo2016}
\APACinsertmetastar {%
guo2016}%
\begin{APACrefauthors}%
Guo, G.%
, Zhu, F.%
, Chen, E.%
, Liu, Q.%
, Wu, L.%
\BCBL {}\ \BBA {} Guan, C.%
\end{APACrefauthors}%
\unskip\
\newblock
\APACrefYearMonthDay{2016}{}{}.
\newblock
{\BBOQ}\APACrefatitle {From footprint to evidence: an exploratory study of
  mining social data for credit scoring} {From footprint to evidence: an
  exploratory study of mining social data for credit scoring}.{\BBCQ}
\newblock
\APACjournalVolNumPages{ACM Transactions on the Web (TWEB)}{10}{4}{1--38}.
\PrintBackRefs{\CurrentBib}

\bibitem [\protect \citeauthoryear {%
Hurley%
\ \BBA {} Adebayo%
}{%
Hurley%
\ \BBA {} Adebayo%
}{%
{\protect \APACyear {2016}}%
}]{%
Hurley2016}
\APACinsertmetastar {%
Hurley2016}%
\begin{APACrefauthors}%
Hurley, M.%
\BCBT {}\ \BBA {} Adebayo, J.%
\end{APACrefauthors}%
\unskip\
\newblock
\APACrefYearMonthDay{2016}{}{}.
\newblock
{\BBOQ}\APACrefatitle {Credit Scoring in the era of Big Data} {Credit scoring
  in the era of big data}.{\BBCQ}
\newblock
\APACjournalVolNumPages{Yale Journal of Law \& Technology}{18}{}{149-216}.
\PrintBackRefs{\CurrentBib}

\bibitem [\protect \citeauthoryear {%
Lawrence%
\ \BBA {} Solomon%
}{%
Lawrence%
\ \BBA {} Solomon%
}{%
{\protect \APACyear {2012}}%
}]{%
Lawrence2012}
\APACinsertmetastar {%
Lawrence2012}%
\begin{APACrefauthors}%
Lawrence, D.%
\BCBT {}\ \BBA {} Solomon, A.%
\end{APACrefauthors}%
\unskip\
\newblock
\APACrefYear{2012}.
\newblock
\APACrefbtitle {{Managing a Consumer Lending Business}} {{Managing a Consumer
  Lending Business}}.
\newblock
\APACaddressPublisher{}{Solomon Lawrence Partners}.
\PrintBackRefs{\CurrentBib}

\bibitem [\protect \citeauthoryear {%
Lessmann%
, Baesens%
, Seow%
\BCBL {}\ \BBA {} Thomas%
}{%
Lessmann%
\ \protect \BOthers {.}}{%
{\protect \APACyear {2015}}%
}]{%
LESSMANN2015124}
\APACinsertmetastar {%
LESSMANN2015124}%
\begin{APACrefauthors}%
Lessmann, S.%
, Baesens, B.%
, Seow, H\BHBI V.%
\BCBL {}\ \BBA {} Thomas, L\BPBI C.%
\end{APACrefauthors}%
\unskip\
\newblock
\APACrefYearMonthDay{2015}{}{}.
\newblock
{\BBOQ}\APACrefatitle {Benchmarking state-of-the-art classification algorithms
  for credit scoring: An update of research} {Benchmarking state-of-the-art
  classification algorithms for credit scoring: An update of research}.{\BBCQ}
\newblock
\APACjournalVolNumPages{European Journal of Operational Research}{247}{1}{124 -
  136}.
\PrintBackRefs{\CurrentBib}

\bibitem [\protect \citeauthoryear {%
Lundberg%
\ \BBA {} Lee%
}{%
Lundberg%
\ \BBA {} Lee%
}{%
{\protect \APACyear {2017}}%
}]{%
shap2017}
\APACinsertmetastar {%
shap2017}%
\begin{APACrefauthors}%
Lundberg, S\BPBI M.%
\BCBT {}\ \BBA {} Lee, S\BHBI I.%
\end{APACrefauthors}%
\unskip\
\newblock
\APACrefYearMonthDay{2017}{}{}.
\newblock
{\BBOQ}\APACrefatitle {A Unified Approach to Interpreting Model Predictions} {A
  unified approach to interpreting model predictions}.{\BBCQ}
\newblock
\BIn{} I.~Guyon\ \BOthers {.}\ (\BEDS), \APACrefbtitle {Advances in Neural
  Information Processing Systems 30} {Advances in neural information processing
  systems 30}\ (\BPGS\ 4765--4774).
\newblock
\APACaddressPublisher{}{Curran Associates, Inc.}
\PrintBackRefs{\CurrentBib}

\bibitem [\protect \citeauthoryear {%
Nayak%
\ \BBA {} Turvey%
}{%
Nayak%
\ \BBA {} Turvey%
}{%
{\protect \APACyear {1997}}%
}]{%
Nayak1997}
\APACinsertmetastar {%
Nayak1997}%
\begin{APACrefauthors}%
Nayak, G\BPBI N.%
\BCBT {}\ \BBA {} Turvey, C\BPBI G.%
\end{APACrefauthors}%
\unskip\
\newblock
\APACrefYearMonthDay{1997}{}{}.
\newblock
{\BBOQ}\APACrefatitle {{Credit Risk Assessment and the Opportunity Costs of
  Loan Misclassification}} {{Credit Risk Assessment and the Opportunity Costs
  of Loan Misclassification}}.{\BBCQ}
\newblock
\APACjournalVolNumPages{Canadian Journal of Agricultural
  Economics}{45}{3}{285--299}.
\PrintBackRefs{\CurrentBib}

\bibitem [\protect \citeauthoryear {%
{\'O}skarsd{\'o}ttir%
, Bravo%
, Sarraute%
, Vanthienen%
\BCBL {}\ \BBA {} Baesens%
}{%
{\'O}skarsd{\'o}ttir%
\ \protect \BOthers {.}}{%
{\protect \APACyear {2019}}%
}]{%
Oskarsdottir2019}
\APACinsertmetastar {%
Oskarsdottir2019}%
\begin{APACrefauthors}%
{\'O}skarsd{\'o}ttir, M.%
, Bravo, C.%
, Sarraute, C.%
, Vanthienen, J.%
\BCBL {}\ \BBA {} Baesens, B.%
\end{APACrefauthors}%
\unskip\
\newblock
\APACrefYearMonthDay{2019}{}{}.
\newblock
{\BBOQ}\APACrefatitle {The Value of Big Data for Credit Scoring: Enhancing
  Financial Inclusion Using Mobile Phone Data and Social Network Analytics}
  {The value of big data for credit scoring: Enhancing financial inclusion
  using mobile phone data and social network analytics}.{\BBCQ}
\newblock
\APACjournalVolNumPages{Applied Soft Computing}{74}{}{26--39}.
\PrintBackRefs{\CurrentBib}

\bibitem [\protect \citeauthoryear {%
Osterwalder%
, Pigneur%
, Smith%
\BCBL {}\ \BBA {} Etiemble%
}{%
Osterwalder%
\ \protect \BOthers {.}}{%
{\protect \APACyear {2020}}%
}]{%
osterwalder2020}
\APACinsertmetastar {%
osterwalder2020}%
\begin{APACrefauthors}%
Osterwalder, A.%
, Pigneur, Y.%
, Smith, A.%
\BCBL {}\ \BBA {} Etiemble, F.%
\end{APACrefauthors}%
\unskip\
\newblock
\APACrefYear{2020}.
\newblock
\APACrefbtitle {The Invincible Company: How to Constantly Reinvent Your
  Organization with Inspiration From the World's Best Business Models} {The
  invincible company: How to constantly reinvent your organization with
  inspiration from the world's best business models}.
\newblock
\APACaddressPublisher{}{Wiley}.
\PrintBackRefs{\CurrentBib}

\bibitem [\protect \citeauthoryear {%
Philippon%
}{%
Philippon%
}{%
{\protect \APACyear {2019}}%
}]{%
Philippon2019}
\APACinsertmetastar {%
Philippon2019}%
\begin{APACrefauthors}%
Philippon, T.%
\end{APACrefauthors}%
\unskip\
\newblock
\APACrefYearMonthDay{2019}{}{}.
\newblock
\APACrefbtitle {On FinTech and Financial Inclusion} {On fintech and financial
  inclusion}\ \APACbVolEdTR{}{\BTR{}\ \BNUM\ NBER Working Paper No. 26330}.
\newblock
\APACaddressInstitution{}{National Bureau of Economic Research}.
\PrintBackRefs{\CurrentBib}

\bibitem [\protect \citeauthoryear {%
Ribeiro%
, Singh%
\BCBL {}\ \BBA {} Guestrin%
}{%
Ribeiro%
\ \protect \BOthers {.}}{%
{\protect \APACyear {2016}}%
}]{%
lime}
\APACinsertmetastar {%
lime}%
\begin{APACrefauthors}%
Ribeiro, M\BPBI T.%
, Singh, S.%
\BCBL {}\ \BBA {} Guestrin, C.%
\end{APACrefauthors}%
\unskip\
\newblock
\APACrefYearMonthDay{2016}{}{}.
\newblock
{\BBOQ}\APACrefatitle {"Why Should {I} Trust You?": Explaining the Predictions
  of Any Classifier} {"why should {I} trust you?": Explaining the predictions
  of any classifier}.{\BBCQ}
\newblock
\BIn{} \APACrefbtitle {Proceedings of the 22nd {ACM} {SIGKDD} International
  Conference on Knowledge Discovery and Data Mining, San Francisco, CA, USA,
  August 13-17, 2016} {Proceedings of the 22nd {ACM} {SIGKDD} international
  conference on knowledge discovery and data mining, san francisco, ca, usa,
  august 13-17, 2016}\ (\BPGS\ 1135--1144).
\PrintBackRefs{\CurrentBib}

\bibitem [\protect \citeauthoryear {%
Salvaire%
}{%
Salvaire%
}{%
{\protect \APACyear {2019}}%
}]{%
Salvaire2019}
\APACinsertmetastar {%
Salvaire2019}%
\begin{APACrefauthors}%
Salvaire, P.%
\end{APACrefauthors}%
\unskip\
\newblock
\APACrefYear{2019}.
\unskip\
\newblock
\APACrefbtitle {Explaining the predictions of a boosted tree algorithm :
  application to credit scoring} {Explaining the predictions of a boosted tree
  algorithm : application to credit scoring}\ \APACtypeAddressSchool
  {\BUMTh}{}{}.
\unskip\
\newblock
\APACaddressSchool {}{Universidade Nova de Lisboa}.
\PrintBackRefs{\CurrentBib}

\bibitem [\protect \citeauthoryear {%
Siddiqi%
}{%
Siddiqi%
}{%
{\protect \APACyear {2017}}%
}]{%
Siddiqi2017}
\APACinsertmetastar {%
Siddiqi2017}%
\begin{APACrefauthors}%
Siddiqi, N.%
\end{APACrefauthors}%
\unskip\
\newblock
\APACrefYear{2017}.
\newblock
\APACrefbtitle {Intelligent Credit Scoring} {Intelligent credit scoring}.
\newblock
\APACaddressPublisher{}{{John Wiley {{\&}} Sons, Inc.}}
\PrintBackRefs{\CurrentBib}

\bibitem [\protect \citeauthoryear {%
Sun%
}{%
Sun%
}{%
{\protect \APACyear {2017}}%
}]{%
Sun2017}
\APACinsertmetastar {%
Sun2017}%
\begin{APACrefauthors}%
Sun, T.%
\end{APACrefauthors}%
\unskip\
\newblock
\APACrefYearMonthDay{2017}{08}{}.
\newblock
{\BBOQ}\APACrefatitle {Balancing Innovation and Risks in Digital Financial
  Inclusion-Experiences of Ant Financial Services Group} {Balancing innovation
  and risks in digital financial inclusion-experiences of ant financial
  services group}.{\BBCQ}
\newblock
\BIn{} (\BPG~37-43).
\PrintBackRefs{\CurrentBib}

\bibitem [\protect \citeauthoryear {%
Sundsøy%
, Bjelland%
, Reme%
, M.Iqbal%
\BCBL {}\ \BBA {} Jahani%
}{%
Sundsøy%
\ \protect \BOthers {.}}{%
{\protect \APACyear {2016}}%
}]{%
Sundsoy2016}
\APACinsertmetastar {%
Sundsoy2016}%
\begin{APACrefauthors}%
Sundsøy, P.%
, Bjelland, J.%
, Reme, B\BHBI A.%
, M.Iqbal, A.%
\BCBL {}\ \BBA {} Jahani, E.%
\end{APACrefauthors}%
\unskip\
\newblock
\APACrefYearMonthDay{2016}{}{}.
\newblock
{\BBOQ}\APACrefatitle {Deep Learning Applied to Mobile Phone Data for
  Individual Income Classification} {Deep learning applied to mobile phone data
  for individual income classification}.{\BBCQ}
\newblock
\BIn{} \APACrefbtitle {2016 International Conference on Artificial
  Intelligence: Technologies and Applications} {2016 international conference
  on artificial intelligence: Technologies and applications}\ (\BPG~96-99).
\newblock
\APACaddressPublisher{}{Atlantis Press}.
\PrintBackRefs{\CurrentBib}

\bibitem [\protect \citeauthoryear {%
Sy%
, Maino%
, Massara%
, Prez-Saiz%
\BCBL {}\ \BBA {} Sharma%
}{%
Sy%
\ \protect \BOthers {.}}{%
{\protect \APACyear {2019}}%
}]{%
Sy2019}
\APACinsertmetastar {%
Sy2019}%
\begin{APACrefauthors}%
Sy, N\BPBI A.%
, Maino, R.%
, Massara, A.%
, Prez-Saiz, H.%
\BCBL {}\ \BBA {} Sharma, P.%
\end{APACrefauthors}%
\unskip\
\newblock
\APACrefYearMonthDay{2019}{}{}.
\newblock
\APACrefbtitle {FinTech in Sub-Saharan African Countries : A Game Changer?}
  {Fintech in sub-saharan african countries : A game changer?}\
  \APACbVolEdTR{}{\BTR{}\ \BNUM\ 19/04}.
\newblock
\APACaddressInstitution{}{International Monetary Fund, African Department}.
\PrintBackRefs{\CurrentBib}

\bibitem [\protect \citeauthoryear {%
{Task Force on Financial Technology}%
}{%
{Task Force on Financial Technology}%
}{%
{\protect \APACyear {2019}}%
}]{%
Wu2019}
\APACinsertmetastar {%
Wu2019}%
\begin{APACrefauthors}%
{Task Force on Financial Technology}.%
\end{APACrefauthors}%
\unskip\
\newblock
\APACrefYearMonthDay{2019}{}{}.
\newblock
\APACrefbtitle {Examining the Use of Alternative Data in Underwriting and
  Credit Scoring to Expand Access to Credit: Hearings before the Task Force on
  Financial Technology.} {Examining the use of alternative data in underwriting
  and credit scoring to expand access to credit: Hearings before the task force
  on financial technology.}
\newblock
\APACrefnote{{US House of Representatives, 116th Cong.}}
\PrintBackRefs{\CurrentBib}

\bibitem [\protect \citeauthoryear {%
Thomas%
, Crook%
\BCBL {}\ \BBA {} Edelman%
}{%
Thomas%
\ \protect \BOthers {.}}{%
{\protect \APACyear {2017}}%
}]{%
thomas2017}
\APACinsertmetastar {%
thomas2017}%
\begin{APACrefauthors}%
Thomas, L.%
, Crook, J.%
\BCBL {}\ \BBA {} Edelman, D.%
\end{APACrefauthors}%
\unskip\
\newblock
\APACrefYear{2017}.
\newblock
\APACrefbtitle {Credit {{Scoring}} and its {Applications}} {Credit {{Scoring}}
  and its {Applications}}\ (\PrintOrdinal{Second Edition}\ \BEd).
\newblock
\APACaddressPublisher{{USA}}{{SIAM}}.
\PrintBackRefs{\CurrentBib}

\bibitem [\protect \citeauthoryear {%
{Valuates Reports}%
}{%
{Valuates Reports}%
}{%
{\protect \APACyear {2019}}%
}]{%
valuatesreports2019}
\APACinsertmetastar {%
valuatesreports2019}%
\begin{APACrefauthors}%
{Valuates Reports}.%
\end{APACrefauthors}%
\unskip\
\newblock
\APACrefYearMonthDay{2019}{{\APACmonth{09}}}{}.
\newblock
\APACrefbtitle {Global {{FinTech Market Size}}, {{Status}} and {{Forecast}}
  2018-2025} {Global {{FinTech Market Size}}, {{Status}} and {{Forecast}}
  2018-2025}\ \APACbVolEdTR{}{\BTR{}\ \BNUM\ QYRE-Othe-2W194}.
\newblock
\APACaddressInstitution{}{Valuates}.
\PrintBackRefs{\CurrentBib}

\bibitem [\protect \citeauthoryear {%
Verbraken%
, Bravo%
, Weber%
\BCBL {}\ \BBA {} Baesens%
}{%
Verbraken%
\ \protect \BOthers {.}}{%
{\protect \APACyear {2014}}%
}]{%
VERBRAKEN2014505}
\APACinsertmetastar {%
VERBRAKEN2014505}%
\begin{APACrefauthors}%
Verbraken, T.%
, Bravo, C.%
, Weber, R.%
\BCBL {}\ \BBA {} Baesens, B.%
\end{APACrefauthors}%
\unskip\
\newblock
\APACrefYearMonthDay{2014}{}{}.
\newblock
{\BBOQ}\APACrefatitle {Development and application of consumer credit scoring
  models using profit-based classification measures} {Development and
  application of consumer credit scoring models using profit-based
  classification measures}.{\BBCQ}
\newblock
\APACjournalVolNumPages{European Journal of Operational Research}{238}{2}{505 -
  513}.
\PrintBackRefs{\CurrentBib}

\bibitem [\protect \citeauthoryear {%
{World Artificial Intelligence Conference}%
}{%
{World Artificial Intelligence Conference}%
}{%
{\protect \APACyear {2019}}%
}]{%
waic2019}
\APACinsertmetastar {%
waic2019}%
\begin{APACrefauthors}%
{World Artificial Intelligence Conference}.%
\end{APACrefauthors}%
\unskip\
\newblock
\APACrefYearMonthDay{2019}{Aug}{}.
\newblock
\APACrefbtitle {Lufax {CTO} {M}ao {J}inliang: {AI} is reshaping the wealth
  management industry.} {Lufax {CTO} {M}ao {J}inliang: {AI} is reshaping the
  wealth management industry.}
\newblock
\APACrefnote{Accessed 2020-04-02}
\PrintBackRefs{\CurrentBib}

\bibitem [\protect \citeauthoryear {%
Xiaa%
, Liu%
, Li%
\BCBL {}\ \BBA {} Liu%
}{%
Xiaa%
\ \protect \BOthers {.}}{%
{\protect \APACyear {2017}}%
}]{%
Xia2017}
\APACinsertmetastar {%
Xia2017}%
\begin{APACrefauthors}%
Xiaa, Y.%
, Liu, C.%
, Li, Y.%
\BCBL {}\ \BBA {} Liu, N.%
\end{APACrefauthors}%
\unskip\
\newblock
\APACrefYearMonthDay{2017}{}{}.
\newblock
{\BBOQ}\APACrefatitle {Boosted decision tree approach using Bayesian
  hyper-parameter optimization for credit scoring} {Boosted decision tree
  approach using bayesian hyper-parameter optimization for credit
  scoring}.{\BBCQ}
\newblock
\APACjournalVolNumPages{Expert Systems with Applications}{78}{}{225--241}.
\PrintBackRefs{\CurrentBib}

\bibitem [\protect \citeauthoryear {%
W\BHBI Y.~Zhang%
}{%
W\BHBI Y.~Zhang%
}{%
{\protect \APACyear {2016}}%
}]{%
Zhang2016}
\APACinsertmetastar {%
Zhang2016}%
\begin{APACrefauthors}%
Zhang, W\BHBI Y.%
\end{APACrefauthors}%
\unskip\
\newblock
\APACrefYearMonthDay{2016}{}{}.
\newblock
{\BBOQ}\APACrefatitle {Exploring the Improved Personal Credit Scoring Model of
  Ant Financial Services in its Disruptive Innovation Process} {Exploring the
  improved personal credit scoring model of ant financial services in its
  disruptive innovation process}.{\BBCQ}
\newblock
\BIn{} \APACrefbtitle {3d International Conference on Applied Social Science
  Research (ICASSR 2015)} {3d international conference on applied social
  science research (icassr 2015)}\ (\BPG~408-410).
\newblock
\APACaddressPublisher{}{Atlantis Press}.
\PrintBackRefs{\CurrentBib}

\bibitem [\protect \citeauthoryear {%
Y.~Zhang%
, Jia%
, Diao%
, Hai%
\BCBL {}\ \BBA {} Li%
}{%
Y.~Zhang%
\ \protect \BOthers {.}}{%
{\protect \APACyear {2016}}%
}]{%
ZhangHengyue2016}
\APACinsertmetastar {%
ZhangHengyue2016}%
\begin{APACrefauthors}%
Zhang, Y.%
, Jia, H.%
, Diao, Y.%
, Hai, M.%
\BCBL {}\ \BBA {} Li, H.%
\end{APACrefauthors}%
\unskip\
\newblock
\APACrefYearMonthDay{2016}{}{}.
\newblock
{\BBOQ}\APACrefatitle {Research on Credit Scoring by Fusing Social Media
  Information in Online Peer-to-Peer Lending} {Research on credit scoring by
  fusing social media information in online peer-to-peer lending}.{\BBCQ}
\newblock
\APACjournalVolNumPages{Procedia Computer Science}{91}{}{168 - 174}.
\newblock
\APACrefnote{Promoting Business Analytics and Quantitative Management of
  Technology: 4th International Conference on Information Technology and
  Quantitative Management (ITQM 2016)}
\PrintBackRefs{\CurrentBib}

\end{thebibliography}

\end{document}